\documentclass[12pt, twoside, here]{article}
\usepackage{epsf}
\usepackage{times,colordvi,amsmath,epsfig,float,color,multicol}
\usepackage{graphics}
\usepackage{hhline}
\usepackage{graphicx}
\usepackage[large]{subfigure}
\usepackage[latin1]{inputenc}
\usepackage{rotating}
\usepackage{marvosym}
\usepackage{amsfonts}
\usepackage{amssymb}

\newtheorem{assumption}{\sc Assumption}

\title{The Finite Element Implementation of a K.P.P. Equation for the Simulation
of Tsetse Control Measures in the Vicinity of a Game Reserve}

\author{S. J. Childs \\ \\ {\small\em ARC-Onderstepoort Veterinary Institute, Private Bag X5,} \\ {\small\em Pretoria, 0110, South Africa.} \\ {\small\em Tel: +27 72 8459556. E-mail: simonjohnchilds@gmail.com}}

\renewcommand{\thefootnote}{\fnsymbol{footnote}}
\date{Mathematical Biosciences, 227: 29--43, 2010}       

\begin{document}

\maketitle
\renewcommand{\thefootnote}{\arabic{footnote}}

\begin{abstract}
\noindent {\em An equation, strongly reminiscent of Fisher's equation, is used
to model the response of tsetse populations to proposed control measures in the
vicinity of a game reserve. The model assumes movement is by diffusion and that
growth is logistic. This logistic growth is dependent on an historical
population, in contrast to Fisher's equation which bases it on the present
population. The model therefore takes into account the fact that new additions
to the adult fly population are, in actual fact, the descendents of a population
which existed one puparial duration ago, furthermore, that this puparial
duration is temperature dependent. Artificially imposed mortality is modelled as
a proportion at a constant rate. Fisher's equation is also solved as a formality. \\

\noindent The temporary imposition of a 2 \% $\mathrm{day}^{-1}$ mortality
everywhere outside the reserve for a period of 2 years will have no lasting
effect on the influence of the reserve on either the Glossina austeni or the G.
brevipalpis populations, although it certainly will eradicate tsetse from poor
habitat, outside the reserve. A 5 $\mathrm{km}$-wide barrier with a minimum
mortality of 4 \% $\mathrm{day}^{-1}$, throughout, will succeed in isolating a
worst-case, G. austeni population and its associated trypanosomiasis from the
surrounding areas. A more optimistic estimate of its mobility suggests a
mortality of 2 \% $\mathrm{day}^{-1}$ will suffice. For a given target-related
mortality, more mobile species are found to be more vulnerable to eradication
than more sedentary species, while the opposite is true for containment.} 
\end{abstract}


\section{Introduction}

Various testse control measures, in the vicinity of a game reserve, are
experimented with in a simulation context. The simulation models migration as
diffusion, assumes growth is logistic and any artificially imposed mortality is
taken to be a proportion at a constant rate. The problem posed is essentially
one of designing counter measures against the influence of a reserve, of a
particular size and geometry, on tsetse population levels outside its confines.
The animals in the reserve are, moreover, considered to be a reservoir of more
lethal strains of trypanosomiasis, as well as having a generally higher
prevalence. The {\em G. brevipalpis} and {\em G. austeni} populations in and
around the Hluhluwe-iMfolozi Game Reserve are the subject of this case study. 
\begin{figure}[H]
    \begin{center}
\includegraphics[width=15.5cm, angle=0, clip = true]{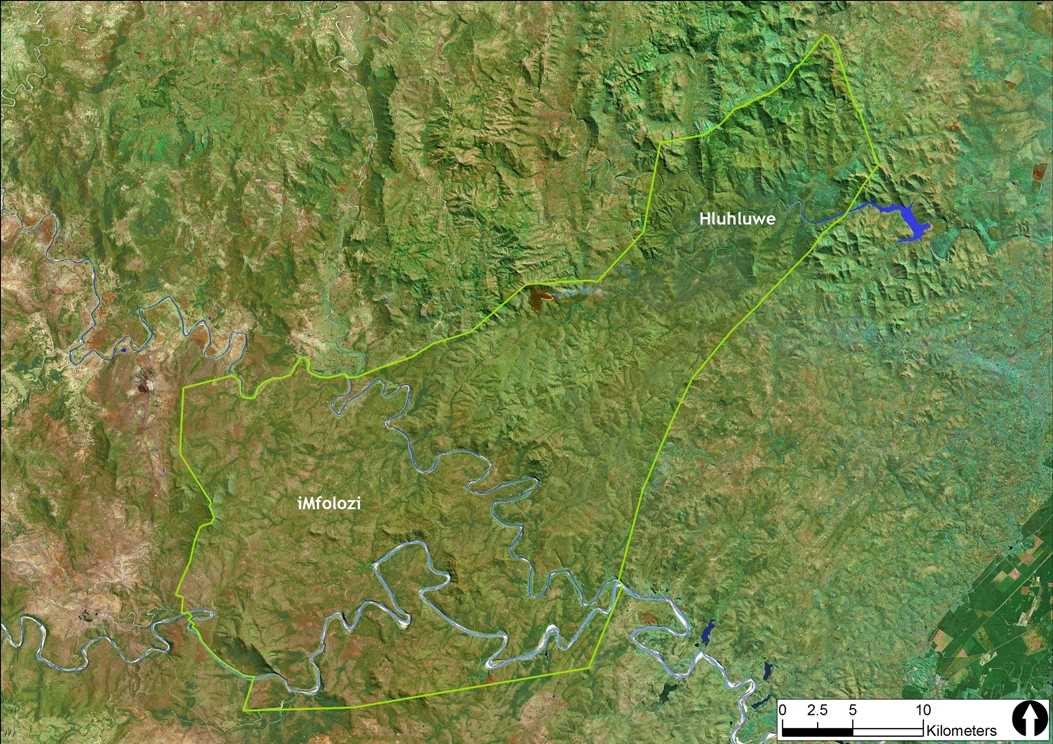}
\caption{Satellite image of the Hluhluwe-iMfolozi Game Reserve and its surroundings.} \label{HluhluweImfoloziSatelliteImage}
   \end{center}
\end{figure}
The Hluhluwe-iMfolozi Game Reserve has the distinction of being the oldest
proclaimed game reserve in Africa. It measures some 960 $\mathrm{km}^2$ and is
located in the southern vicinity of 28 $^\circ \mathrm{S}$ and 32 $^\circ
\mathrm{E}$, in KwaZulu-Natal, South Africa. Inland of the coastal plain and
set in the foothills of the escarpment, the temperature of the region is
somewhat elevated for its latitutde. Riverine forest and thicket make the
reserve the habitat of both {\em G. brevipalpis} and {\em G. austeni}. It is
noteworthy, with regard to both species, that the Hluhluwe River has a flood
plain within the reserve and that the backwater of the Hluhluwe Dam also extends
well into it. At around this position, the reserve is approximately only 25
$\mathrm{km}$  from the St. Lucia estuary; lush habitat and a world heritage
site which lies to the east. Habitat outside the reserve is otherwise degraded
to the extent that the boundary of the reserve is discernable in satellite
images. The presence of tsetse in association with large populations of buffalo
and other wild animals, lead the Hluhluwe-iMfolozi Game Reserve to be a thorn
in the side of neighbouring agriculture. Any tsetse control within the reserve
confines is, however, considered decidedly unwelcome nowadays, despite the
reserve having its origins in an experimental area for the control of {\em G.
pallidipes}\footnotemark[1]\footnotetext[1]{The reserve falls within what was
once the very extensive habitat of {\em G. pallidipes},  a species completely
eradicated from KwaZulu-Natal in the first half of the 20th century.}. 

Much has recently come to light on the vector competence of {\em G. brevipalpis}
and {\em G. austeni}. The predominant infection is that of {\em Trypanasoma
congolense}, {\em T. vivax} being prevalent to a far lesser extent. It is
noteworthy that out of 900 {\em G. brevipalpis} tenerals split into 3 equal
groups and respectively allowed to feed on a different parasitaemic animal, the
midgut of 4 \% and the preboscis of 0 \% were found to be infected ({\sc
Motloang}, {\sc Masumu}, {\sc Van Den Bossche}, {\sc Majiwa} and {\sc Latif}
\cite{MotloangMasumuVanDenBosscheMajiwaLatif}). The prevalence of preboscal
infection for the same experiment involving {\em G. austeni} tenerals was, in
contrast, 12 \% and 19 \% were found to have an infected midgut ({\sc Motloang
et al.} \cite{MotloangMasumuVanDenBosscheMajiwaLatif}). The same authors also
conducted a second experiment in which they challenged each of 7 susceptable
bovines with a different {\em G. brevipalpis} catch, taken from the wild, in an
insect-proof facility (the combined catches totalling 468 specimens). No
infection resulted. The same trial was then conducted by challenging each of 2
bovines and 1 goat with a different {\em G. austeni} catch taken from the wild
(the combined catches totalling a mere 43 specimens). All three challenges
resulted in infection. {\em G. austeni} is therefore a highly competent vector,
while both prevalence and transmission rates are virtually non-existant in the
case of {\em G. brevipalpis}. The issue of mechanical infection by {\em G.
brevipalpis} is currently under investigation by the same authors.

{\sc Williams}, {\sc Dransfield} and {\sc Brightwell} \cite{Williams2}
originally entertained the idea of using Fisher's equation to model the
distribution of tsetse populations and {\sc Hargrove} \cite{Hargrove6} devised
the best implementation his circumstances permitted. The model entertained in
this work is based on a very similar equation and differs mostly in the exact
specification of population density in the logistic part. It belongs to a more
general category of partial differential equations known as
Kolmogoroff-Petrovsky-Piscounoff (K.P.P.) equations. Such partial differential
equations also happen to be parabolic. In this regard, it is important to note
that one cannot simply solve a parabolic, partial differential equation using an
explicit method, nor should one apply the standard finite difference method to
non-rectangular domains. The former is widely accepted as a faux pas and even
in the event of circumstances which favour a correct solution, it has no
credibility whatsoever. As such, the problem is ideally suited to the
application of the finite element method in an implementation which is fully
implicit in time. Fisher's equation is both parabolic and nonlinear.

In the present model flies are assumed to migrate by some kind of Brownian
motion, down a diffusion gradient, based on the random nature of their movement
(observed by {\sc Bursell} \cite{Bursell4} and demonstrated by {\sc Rogers}
\cite{Rogers1}). Growth in the present fly population has its origins in an
historical fly population; one which existed one puparial duration ago. This
puparial duration is temperature dependent. Any artificially imposed mortality
is modelled as a straight-forward proportion at a constant rate. The model
itself exceeded 5000 lines of fairly extensively commented Fortran, while the
mesh generator exceeded 1700 lines. 

Fisher's equation was also solved as a formality and from a point of view of
academic interest. Fisher's equation is not perfectly suited to tsetse
application owing to the large puparial duration which characterises the {\em
Glossina} genus. A logistic term dependent on the present population density is
something known to be incorrect when, in actual fact, the larval deposition,
responsible for growth in the present population, took place a significant time
previously. In the Fisher's-equation model the existance of the pupal phase is
denied, alternatively, pupae are assumed to migrate and reproduce.

\section{Derivation of a Model}

The aim of the model is to predict how a tsetse population becomes distributed
in space and how this distribution changes over time, through migration,
self-regulating growth and artificially imposed mortality. The intention is to
predict a population density, $\rho(\mathbf{x},t)$ (in which $\mathbf{x}$ and
$t$ are space and time respectively), based on these phenomena. 

Which population? The subject of the intended model is the vector of
trypanosomiasis, namely adult tsetse flies. Pupae neither migrate, nor do they
(or any flies belonging to the pre-ovulatory cohort, for that matter) form any
part of the actively reproductive population. For these reasons the population
density, $\rho(\mathbf{x},t)$, is defined not to include pupae. While it is
tempting to also exclude any flies belonging to the pre-ovulatory cohorts from a
reproductive point of view, such flies are mobile and subject to the external,
artificial mortality to be imposed; indeed, the subject of this investigation.
While the correspondence of the reproductive population to the mobile and
vulnerable population is not perfect, it is suitably close.

With the relevant population identified, how might one model its change brought
about through migration, self-regulating growth and artificially imposed
mortality? If all three dynamics can be regarded as being mutually independent
of one another, they can be considered in isolation.

\subsubsection*{Migration}

{\sc Bursell} \cite{Bursell4} put forward the theory that the movement of tsetse
was of a random nature, not unlike Brownian motion, and {\sc Rogers}
\cite{Rogers1} proved these assertions quantitatively. If one can conceive of a
gas as a continuum, it is only slightly more abstract to conceive of a fly
population as a continuum. Consider the hypothetical scenario of a mobile
population in the absence of either reproduction or mortality. Biomass should
therefore be conserved and the standard continuum-mechanical result for mass
conservation pertains. It can be manipulated to give a result not unlike the
Reynolds transport theorem and Fick's first law applied. (A full exposition is
provided in the addendum.) The resulting rate for the effect of migration, in
isolation, is
\begin{eqnarray*}
\frac{\partial \rho}{\partial t} &=& \lambda \ \mathop{\rm div} \nabla \rho
\end{eqnarray*} 
in which $\lambda$ is the diffusion coefficient.

\subsubsection*{Self-Regulating Growth} 

The logistic model needs little introduction to a biological audience. The
population is assumed to grow at a rate which is some proportion, $r$, of the
parent population, $\rho^*$, and this growth rate must necessarily also be
constrained by the carrying capacity, $K$, of the environment. The logistic rate
for self-regulating growth, in isolation, is 
\begin{eqnarray*}
\frac{\partial \rho}{\partial t} &=& r \rho^* \left( 1 - \frac{\rho^*}{K} \right).
\end{eqnarray*} 
Which  is the relevant population? Larval production is clearly dependent on the
parental population which existed one puparial duration ago in the tsetse
context. What about the population density in the second factor of the logistic
term; the one limiting the growth rate? The pertinent population is not as
obvious in this instance. Combined pupal and teneral mortality is an order of
magnitude higher than adult mortality ({\sc Hargrove} \cite{Hargrove3}) and {\sc
Vale} \cite{Vale1} seems to think that parasitism alone accounts for between 40
\% to 60 \% of the overall pupal mortality, under usual circumstances and in a
favourable environment. Quantitative work linking predation and parasitism to
the density at pupal sites has been carried out by {\sc Rogers} and {\sc
Randolph} \cite{RogersAndRandolph1}. That work could therefore be taken to
recommend a logistic term based entirely on an historical population density,
that which existed at the time of parturition.  

Can such a model be reconciled with the other, remaining causes of pupal
mortality? Although fat loss\footnotemark[1] and water
loss\footnotemark[1]\footnotetext[1]{Teneral mortality from both fat loss and
water loss is thought to be high and is often the cumulative effect of
temperature and humidity conditions which prevailed during the pupal phase.} are
determined by the external variables of temperature and humidity, an indirect
dependence on population density is possible in the event of a shortage of
available breeding sites. The spatial variation of temperature and humidity are
otherwise incorporated in the carrying capacity and growth rate of each
environment. It is important to note, however, that any temporal variation in
the growth rate is beyond the scope of the standard logistic model, although
such a model does allow for a time-dependent carrying capacity. 

The fact that an historical population level was responsible for both larval
production and subsequent, density-dependent mortality is taken into account in
this particular model. Both larval production and subsequent natural mortality
are assumed dependent on the historic population level, that which existed one
puparial duration ago. The population density at the time of larval deposition
was
\begin{eqnarray*}
\rho^*(\mathbf{x},t) \equiv \rho(\mathbf{x},t - {\bar \tau}),
\end{eqnarray*}
in which ${\bar \tau}$ is the relevant puparial duration. 

\subsubsection*{Artificially Imposed Mortality} 

Suppose that the effect of targets, pour-ons etc. is to cause the population density to decline according to $\rho \delta t$, where $\delta$ is independent of the population density. Then the resulting rate for an artificially imposed mortality, in isolation, is
\begin{eqnarray*}
\frac{\partial \rho}{\partial t} &=& - \delta \rho.
\end{eqnarray*} 

\subsection{A Governing Equation} 

The combined effect of all three phenomena is additive and a model can therefore
be based on the following equation. Two alternatives arise based on the exact
specification of the parent population density, $\rho^*(\mathbf{x},t)$. In the
equation 
\begin{eqnarray} \label{1}
\frac{\partial \rho(\mathbf{x},t)}{\partial t} &=& \lambda \mathop{\rm div}
\nabla \rho(\mathbf{x},t) + r \rho^*(\mathbf{x},t) \left( 1 -
\frac{\rho^*(\mathbf{x},t)}{K}  \right) - \delta \rho(\mathbf{x},t),
\end{eqnarray}
$\rho(\mathbf{x},t)$ is otherwise the current population density (in which
$\mathbf{x}$ and $t$ are space and time respectively), $\lambda$ is a diffusion
rate, $r$ is the population growth rate, $K$ is the carrying capacity of the
environment and $\delta$ is an artificially imposed mortality. The quantity
$\rho^*(\mathbf{x},t)$ is either an historical population density or the current
population density, depending on the model preferred.

{\sc Remark:} Notice that in the special case of $\rho^*(\mathbf{x},t) = \rho(\mathbf{x},t)$ and $\delta = 0$, Equation \ref{1} becomes immediately recognizable as Fisher's equation in its classical form. It is otherwise part of a more general and widely inclusive family, known as Kolmogoroff-Petrovsky-Piscounoff equations.

\subsubsection*{Limitation}

The correspondence of the reproductive population to the mobile and vulnerable
population is not perfect. The modelled population includes pre-ovulatory flies
which have a slightly longer interlarval period. These pre-ovulatory flies are not
breeding, yet they are involved in logistic growth. An assumption implicit in
the logistic model is therefore a fixed age profile.  
\begin{assumption} \label{assumption1} {\bf \em The age profile of the population is fixed.}
\end{assumption}  
How reasonable is this assumption? One consequence of any artificially imposed,
adult-selective mortality (such as is contemplated) is that a smaller
proportion of reproductive adults should exist than the model supposes. This
gives rise to a damped logistic response from the model at population levels
above $K/2$ and an over-reactive one for population levels below $K/2$. 

\section{Implementation} 

For the purposes of an implementation, Equation \ref{1} can be rewritten in dimensionless form and the resulting primitive variable formulation converted to a variational one,
\begin{eqnarray*} 
\int_{ \Omega } w \ \frac{\partial \rho}{\partial t} \ {d{\Omega}} + \frac{\lambda}{\lambda_{\scriptsize scale}} \int_{ \Omega } {\nabla} w \cdot {{\nabla} {\rho }} \ {d {\Omega}} &=& \frac{r}{r_{\scriptsize scale}} \int_{
\Omega } w \ \rho^* \left( 1 - \frac{ \rho^* }{K} \right) \ {d {\Omega}} -
\frac{ \delta }{r_{\scriptsize scale}} \int_{ \Omega } w \ \rho \ {d {\Omega}}
\end{eqnarray*}
(a full exposition is provided in the addendum). A fully implicit, backward
difference is used for the temporal discretisation, while the finite element
method is used for the spatial discretisation. 

The solution at time $t$ is accordingly assumed to be a linear combination of
shape functions, ${\mathbf{\psi}}({\mathbf{x}})$. That is,
\[ 
\rho({\mathbf{x}}) {\mid}_t = \sum_{i=1}^{N}
{c}_i \psi_i({\mathbf{x}}),
\]
where the $c_i$ are the constants of the finite element approximation (the nodal
solutions) and $N$ is the total number of nodes. The problem on each element is
calculated in terms of a standard, master element coordinate system, $\{
{\mathbf{\xi}} \}$. The approximate equation, to be solved for the nodal
population densities, $P^e_j$ (pertaining to element $e$), is consequently
\begin{eqnarray*}
&& \hspace{-25mm} {\mathop{\mbox{\LARGE\bf\sf A}}}_{e=1}^E \ \left\{ \ \frac{1}{\Delta t}
\int_{\hat{\Omega}} \phi_{i} \phi_{j} J^e {d{\hat{\Omega}}} \ + \
\frac{\lambda}{\lambda_{\scriptsize \mbox{scale}}} \ \int_{\hat{\Omega}} 
\frac{\partial \phi_{i}}{\partial x_k} \frac{\partial \phi_{j}}{\partial x_k}
J^e {d{\hat{\Omega}}} \ + \ \frac{ \delta }{ r_{\scriptsize \mbox{scale}} } \int_{\hat{\Omega}} \phi_{i} \phi_{j} J^e {d{\hat{\Omega}}} \right\} {\mathop{\mbox{\LARGE\bf\sf A}}}_{e=1}^E \ P^e_j
\hspace{0mm} \nonumber \\ 
\hspace{10mm} &=& \ {\mathop{\mbox{\LARGE\bf\sf A}}}_{e=1}^E \ \left\{
\frac{1}{{\Delta}t} \int_{\hat{\Omega}} \phi_{i} \phi_{m} J^e {d{\hat{\Omega}}}
\ P^e_m{\mid}_{t - {\Delta}t} \ + \ \frac{r}{r_{\scriptsize \mbox{scale}}} \
\int_{\hat{\Omega}} \phi_{i} \phi_{n} J^e {d{\hat{\Omega}}} \ P_n^{e}{\mid}_{t -
{\bar \tau}} \right. \nonumber \\ 
\hspace{10mm} && - \  \frac{r}{r_{\scriptsize \mbox{scale}}} \ \left.
\int_{\hat{\Omega}} \phi_{i} \phi_{l} \frac{ \phi_{j} }{ K } J^e
{d{\hat{\Omega}}} \ P_l^{e}{\mid}_{t - {\bar \tau}} \  \
P_j^{e}{\mid}_{t - {\bar \tau}} \ \right\},  
\end{eqnarray*}
in which ${\mathop{\mbox{\LARGE\bf\sf A}}}$ is the element assembly
operator, $E$ is the total number of elements, $e$, into which the
domain has been subdivided, $\hat{\Omega}$ is the master element
domain, $\Delta t$ is the length of the time step, the
$\phi_i(\mathbf{\xi})$  are the basis functions,
\renewcommand{\thefootnote}{\fnsymbol{footnote}}
\begin{eqnarray*} 
\frac{\partial \phi_{i}}{\partial x_j}( {\mathbf{\xi}} ) = \frac{\partial \phi_{i}}{\partial{\xi}_k}
\frac{{\xi}_k}{\partial x_j} \ , \hspace{10mm} {J^e} = \left| \frac{\partial {\mathbf{x}}}{\partial \mbox{\boldmath{$\xi$}} } \right| \ \mbox{for element e},
\end{eqnarray*}
$\lambda$ is the rate of diffusion, $\lambda_{\scriptsize \mbox{scale}}$ is a
diffusion rate scale, $r$ is the population growth rate, $r_{\scriptsize
\mbox{scale}}$ is a population growth rate scale, $K$ is the carrying capacity
of the environment and $\delta$ is an artificially imposed mortality.
$P_n^{e}{\mid}_{t - {\bar \tau}}$ denotes the solution at the time of larval
deposition (that which led to the present eclosion), ${\bar \tau}$ being the
relevant average of puparial durations. The second order accurate linearisation originally used in {\sc Childs} \cite{Childs1},
\[
2 \mathbf{P}^e \mid_{t - \Delta t} - \mathbf{P}^e \mid_{t - 2 \Delta t},
\]
was used for the first iteration of the nonlinear term arising in the analogous implementation of Fisher's equation.
\renewcommand{\thefootnote}{\arabic{footnote}}

\section{The Relevant Parental Population}

The relevant parental population is that which existed one puparial duration
ago. Determining the puparial duration leading to the present eclosion is a minor problem in its own right. It is known that at a given temperature, $T$, the puparial duration, in days, can be calculated according to the formula
\begin{eqnarray} \label{2}
\tau(T) &=& \frac{ 1 + e^{a + bT} }{\kappa},
\end{eqnarray} 
({\sc Phelps and Burrows} \cite{phelpsAndBurrows1}). For females, $\kappa =
0.057 \pm 0.001$, $a = 5.5 \pm 0.2$ and $b = -0.25 \pm 0.01$. For males, $\kappa
= 0.053 \pm 0.001$, $a = 5.3 \pm 0.2$ and $b = -0.24 \pm 0.01$ ({\sc Hargrove}
\cite{Hargrove3}). The puparial durations of all species, with the exception of
{\em G. brevipalpis}, are thought to lie within 10\% of the value predicted by
this formula ({\sc Parker} \cite{Parker1}). {\em G. brevipalpis} takes a little
longer. The shortest puparial duration is that of {\em G. austeni}.

If ${\bar \tau}$ is the relevant average of puparial durations (which is, of course, dependent on itself) then
\begin{eqnarray*} 
{\bar \tau} &\equiv& \frac{1}{\bar \tau} \left[ \left[ \frac{}{} t - \mbox{floor}\left\{ t \right\} \right] \tau(T_{\scriptsize \mbox{ceil}\left\{ t \right\}}) \ + \sum_{i=\mbox{\scriptsize floor}\left\{ t \right\}}^{ \mbox{\scriptsize ceil}\left\{ t - {\bar \tau} + 1 \right\} } \tau(T_{\mbox{\scriptsize }{\scriptsize i}}) \right. \nonumber \\ 
&& \hspace{52mm} + \ \left. \left[ \frac{}{} \mbox{ceil}\left\{ t - {\bar \tau} \right\} - (t - {\bar \tau}) \right] \ \tau(T_{{\mbox{\scriptsize ceil}} \left\{ t - {\bar \tau} \right\}}) \displaystyle \frac{}{} \right],
\end{eqnarray*} 
in which $\tau(T)$ is given by the formula Equation \ref{2} and $T$ is the mean
daily temperature on the day indicated by the subscript. Newton's method is used
in solving the above equation. The relevant parental population at the time $t -
{\bar \tau}$ is a weighted average of the nearest two solutions since a backward
difference was used for the temporal discretisation.

\section{Application of the Model to Hluhluwe-iMfolozi}

{\em G. brevipalpis} and {\em G. austeni} are, in all likelihood, not the most
suitable species for the application of such a model. This is since both forest
species are thought to be fairly specialised and habitat-specific. This
observation is independently born out by the {\sc Rogers} and {\sc Robinson}
\cite{RogersAndRobinson} study (based on {\sc Ford} and {\sc Katondo}
\cite{FordAndKatondo}'s maps) as well as the pupal water loss model in {\sc
Childs} \cite{Childs2}. {\em G. brevipalpis} would, more generally, appear to be
regionally associated with the riverine forest, or thicket, adjacent to drainage
lines. While its pupal habitat appears to be more stringently confined than that
of {\em G. austeni} ({\sc Childs} \cite{Childs2}), the present work will suggest
{\em G. brevipalpis} to be more far-ranging. {\em G. austeni} would, in
contrast, appear to be relatively sedentary and less restricted by habitat.
Nothing appears to be known about the diffusion rates of either {\em G.
brevipalpis} or {\em G. austeni} and so-called worst-case values must be
assumed. {\sc Rogers}' \cite{Rogers1} experiments with {\em G. fuscipes
fuscipes} were in fairly uniform habitat and even then there was
light-sensitive preference. Little of relevance is otherwise known of {\em G.
brevipalpis} and {\em G. austeni}. Their puparial durations subscribe the worst
to the formula Equation \ref{2} ({\sc Parker} \cite{Parker1}). {\em G.
austeni}'s puparial duration is the shortest and {\em G. brevipalpis}' is, by
far, the longest. One would certainly prefer to be modelling savannah species. A
strong affinity to habitat does, nonetheless, present a certain opportunity in
ellucidating diffusion rates, as will presently be demonstrated. 

Pertinent carrying capacities, growth rates and diffusion coefficients need to
be associated with the nodes of the finite element mesh. The initial
Hluhluwe-iMfolozi case study does not attempt to mimic reality to exactness.
Instead, it is rather simplistic and more humble than the competancy of the
model itself allows. 

\subsection{The Carrying Capacity, $K$}

The {\em Glossina} genus is a $K$-strategist and carrying capacities are
therefore important. Figure \ref{risk} is primarily concerned with distribution.
The suggestion is, nevertheless, that a tsetse haven with a zone of influence is
a premise on which to proceed. The {\em G. brevipalpis} data are certainly
indicative of such a reality. The probability of {\em G. austeni} occurring
within the reserve reaches a maximum of around 0.75 and one might speculate a
smaller range due to the small size of this fly. The influence of minor habitat
existing to the east of the reserve might therefore be disregarded on such a
basis.
 
\begin{table}[H]
    \begin{center}
\begin{tabular}{c r | c c c c c}  
&  &  &  &  & \\
& $\approx d$ / $\mathrm{km}$  \hspace{5mm} & \hspace{5mm} $d \le 0$ & \hspace{5mm} $0 < d \le 2.5$ & \hspace{5mm} $2.5 < d \le 5$& \hspace{5mm}$d > 5$ \\ 
&  &  &  &  & \\ \hline 
&  &  &  &  & \\
& {\em G. brevipalpis} \hspace{5mm} & \hspace{5mm} 100 & \hspace{5mm} 20 & \hspace{5mm} 10 & \hspace{5mm} 5 \\ 
$K$ / \% \hspace{5mm} &  &  &  & \\ 
& {\em G. austeni} \hspace{5mm} & \hspace{5mm} 75 & \hspace{5mm} 20 & \hspace{5mm} 10 & \hspace{5mm} 5 \\ 
&  &  &  &  & \\ 
\end{tabular}
\caption{Modelled {\em G. brevipalpis} and {\em G. austeni} carrying capacities designated according to the approximate distance, $d$, from the reserve boundary.} \label{carryingCapacity}
    \end{center}
\end{table}
A lack of data and the simplicity of an initial case study were deemed to
vindicate such a simplistic approach. Note that it includes the assumption that
$K$ is constant in time.    

\subsection{The Maximum Growth Rate, $r$}

At Hluhluwe-iMfolozi mean annual temperatures, each female would produce four
pupae based on {\sc Glasgow} \cite{Glasgow1}'s 49-day, average, adult
life-span. The population would therefore grow by 2.8 \% \Female $^{-1}$
$\mathrm{day}^{-1}$ in the absence of any early mortality, assuming an equal
ratio of the sexes. Of course, in the real world there is massive pupal and
teneral mortality. In reality, the growth rate is probably closer to 0.85 \%
$\mathrm{day}^{-1}$ ({\sc Hargrove} \cite{Hargrove3}). A logistic growth rate of
1.7 \% $\mathrm{day}^{-1}$ was used for good habitat. One of the limitations of
a logistic model is that $r$ is constant in time.
\begin{table}[H]
    \begin{center}
\begin{tabular}{c | c c c c}  
&  &  &  & \\
$\approx d$ / $\mathrm{km}$  \hspace{5mm} & \hspace{5mm} $d \le 0$ & \hspace{5mm} $0 < d \le 2.5$ & \hspace{5mm} $2.5 < d \le 5$& \hspace{5mm}$d > 5$ \\ 
&  &  &  & \\ \hline 
&  &  &  & \\
$r$ / \% $\mathrm{day}^{-1}$ \hspace{5mm} & \hspace{5mm} 1.7 & \hspace{5mm} 0.5 & \hspace{5mm} 0.2 & \hspace{5mm} 0.1 \\ 
&  &  &  & \\ 
\end{tabular}
\caption{The modelled growth rate designated according to the approximate distance, $d$, from the reserve boundary.} \label{growthRate}
    \end{center}
\end{table}

\subsection{The Diffusion Coefficient, $\lambda$} \label{diffusion}

\begin{figure}
    \begin{center}
\includegraphics[height=9.8cm, angle=0, clip = true]{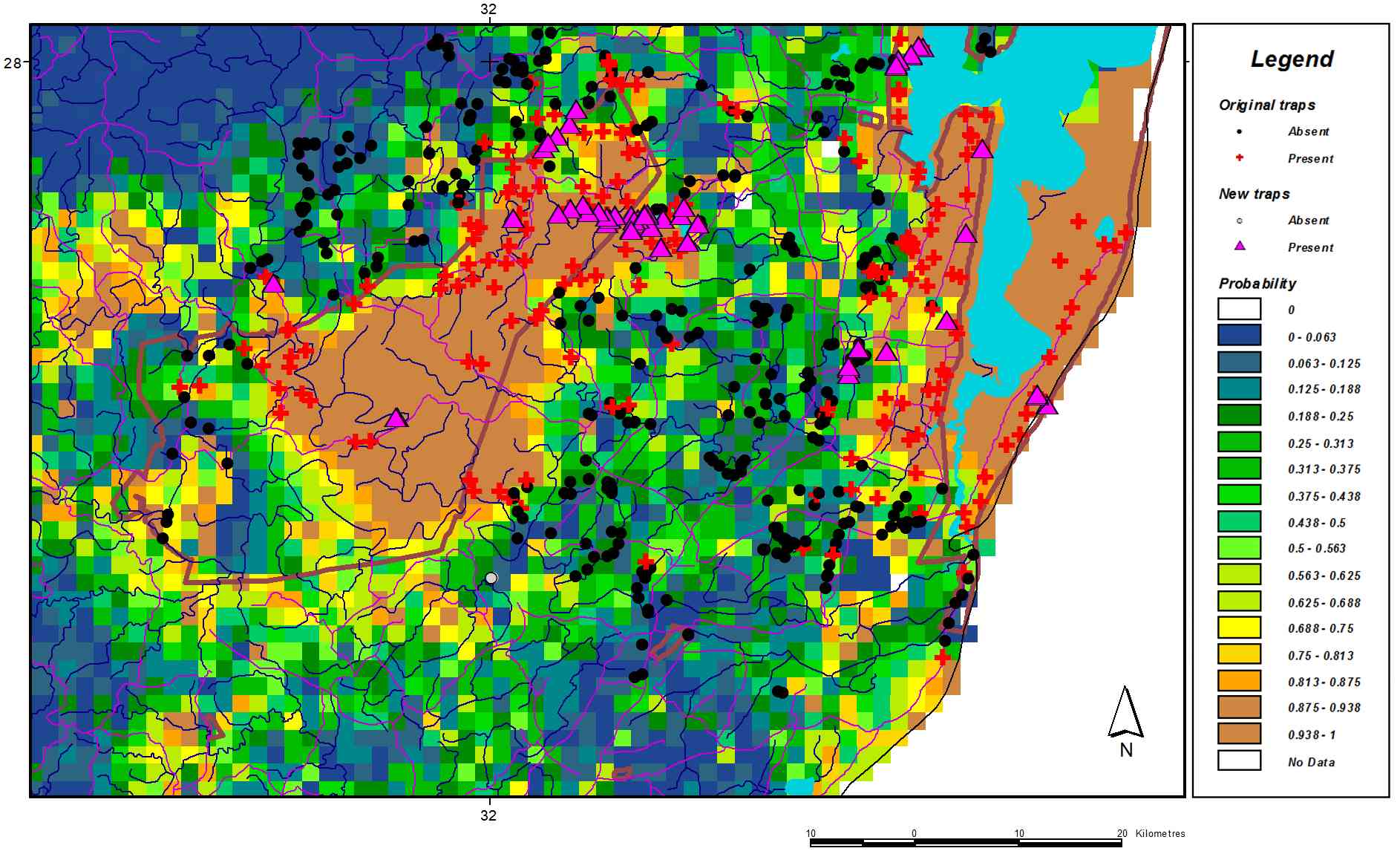}
\includegraphics[height=9.9cm, angle=0, clip = true]{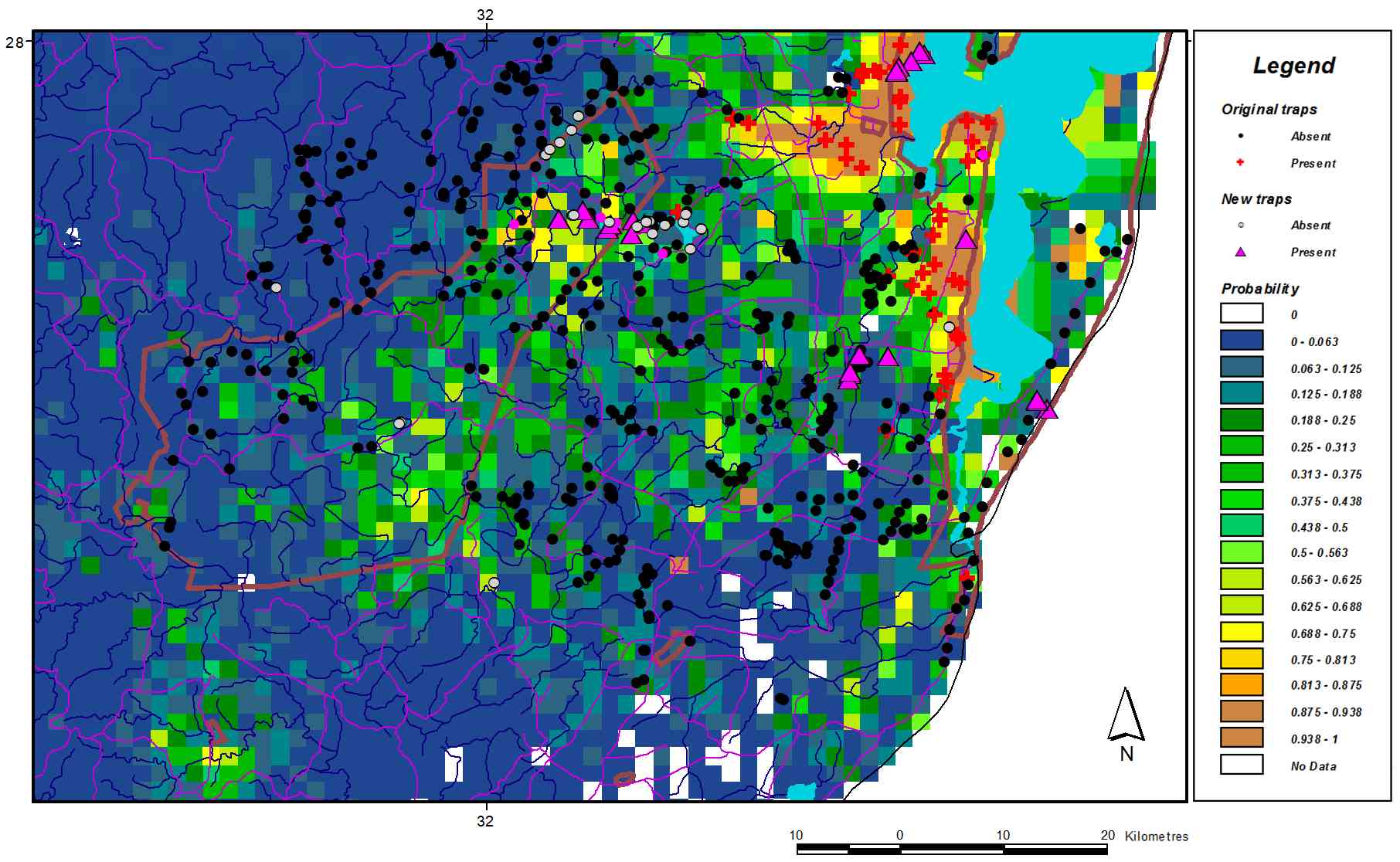}
\caption{{\em G. brevipalpis} risk (top) and {\em G. austeni} risk (bottom). Source: {\sc Hendrickx} \cite{Hendrickx}.} \label{risk}
   \end{center}
\end{figure}
\begin{figure}
\begin{center}
{\bf Candidates for the Worst-Case Diffusion Rates of {\em G. austeni} and {\em G. brevipalpis}} 
\end{center}
\vspace{10mm}
    \begin{center}
\includegraphics[width=7.7cm, angle=0, clip = true]{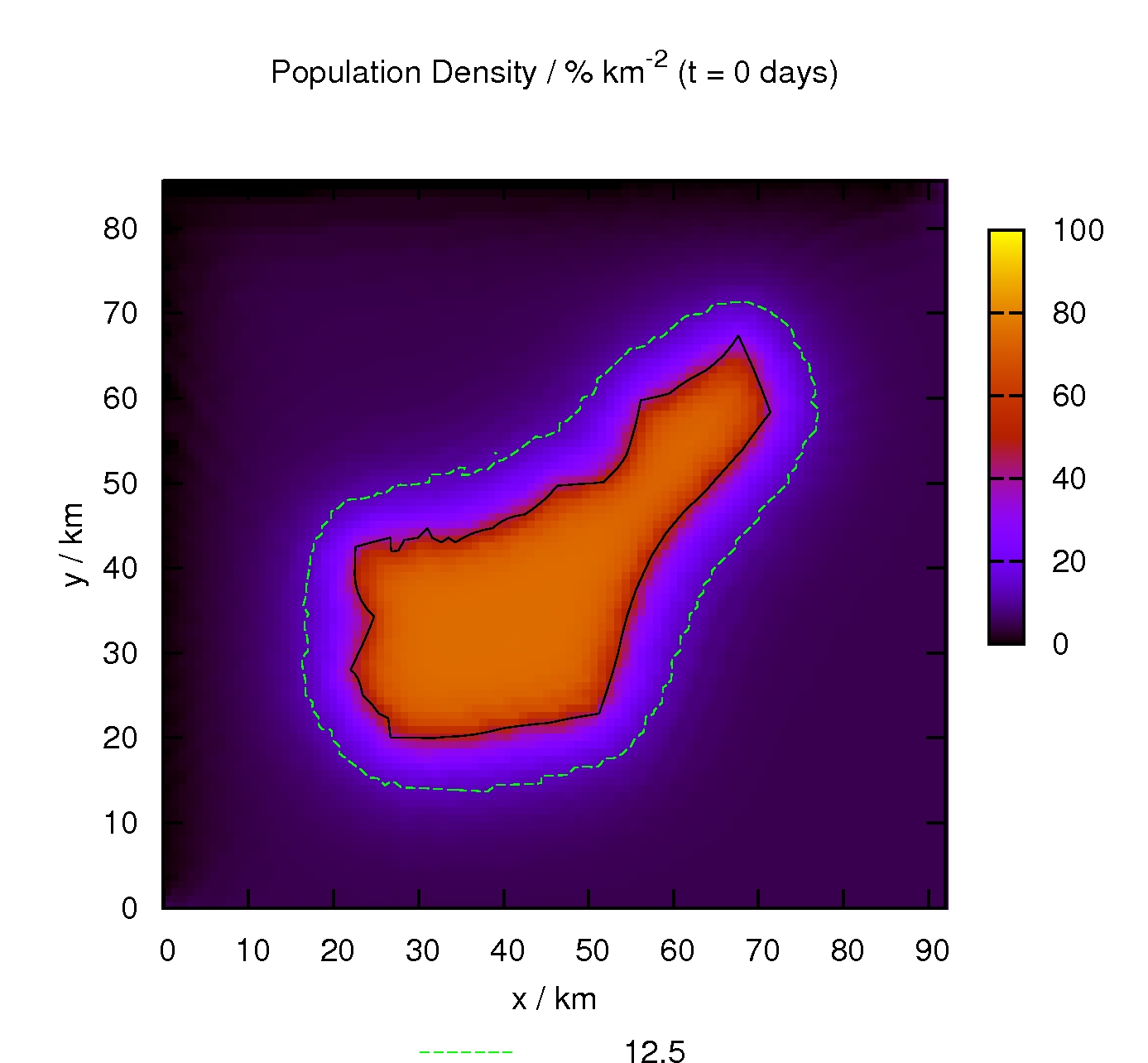}
\includegraphics[width=7.7cm, angle=0, clip = true]{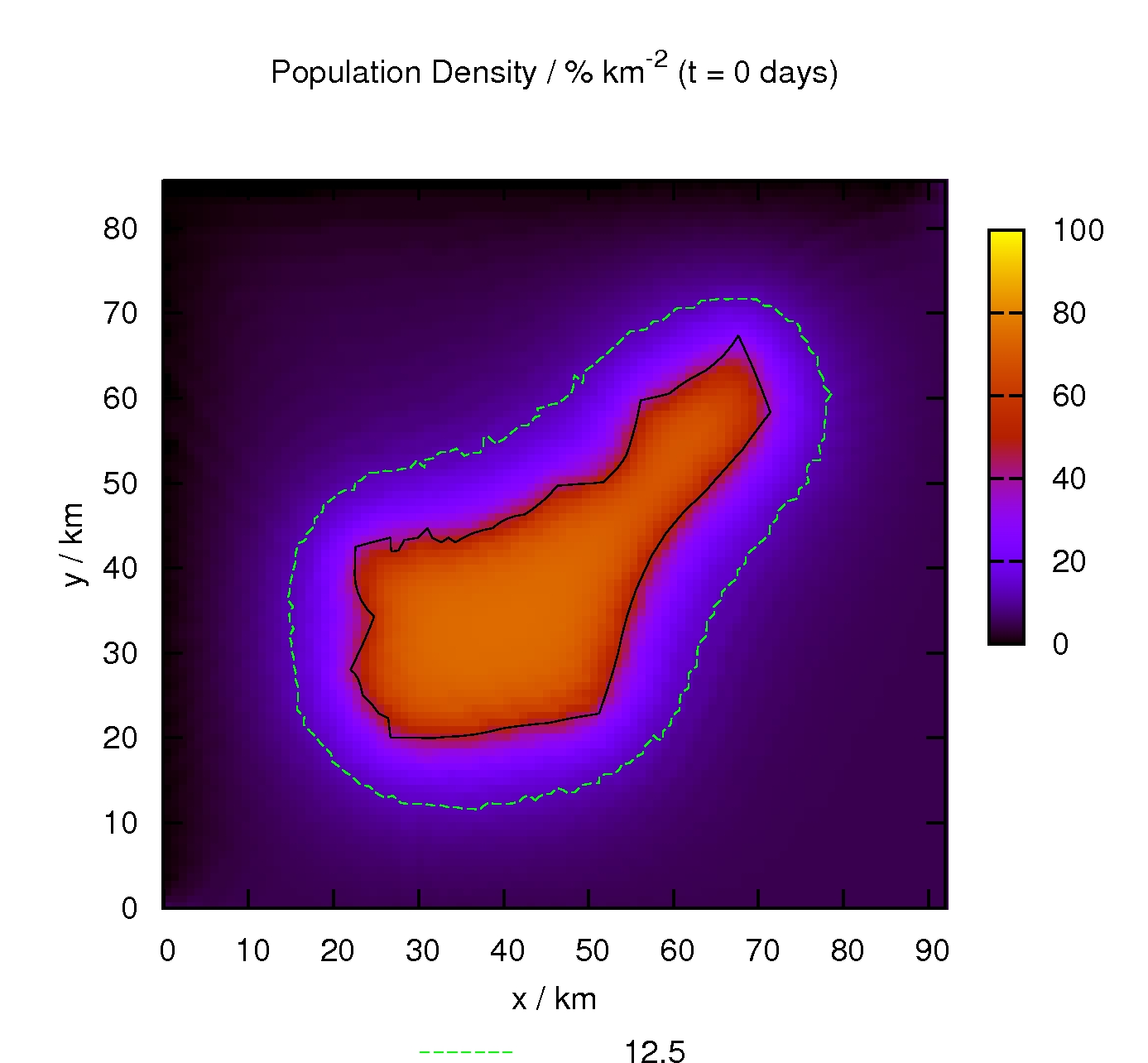}
\caption{Computed steady states for $\lambda = 0.04 \ \mathrm{km}^2 \
\mathrm{day}^{-1}$ (at left) and $\lambda = 0.08 \ \mathrm{km}^2 \
\mathrm{day}^{-1}$ (at right) using {\em G. austeni} carrying capacities.}
\label{1st}
\vspace{5mm}
\includegraphics[width=7.7cm, angle=0, clip = true]{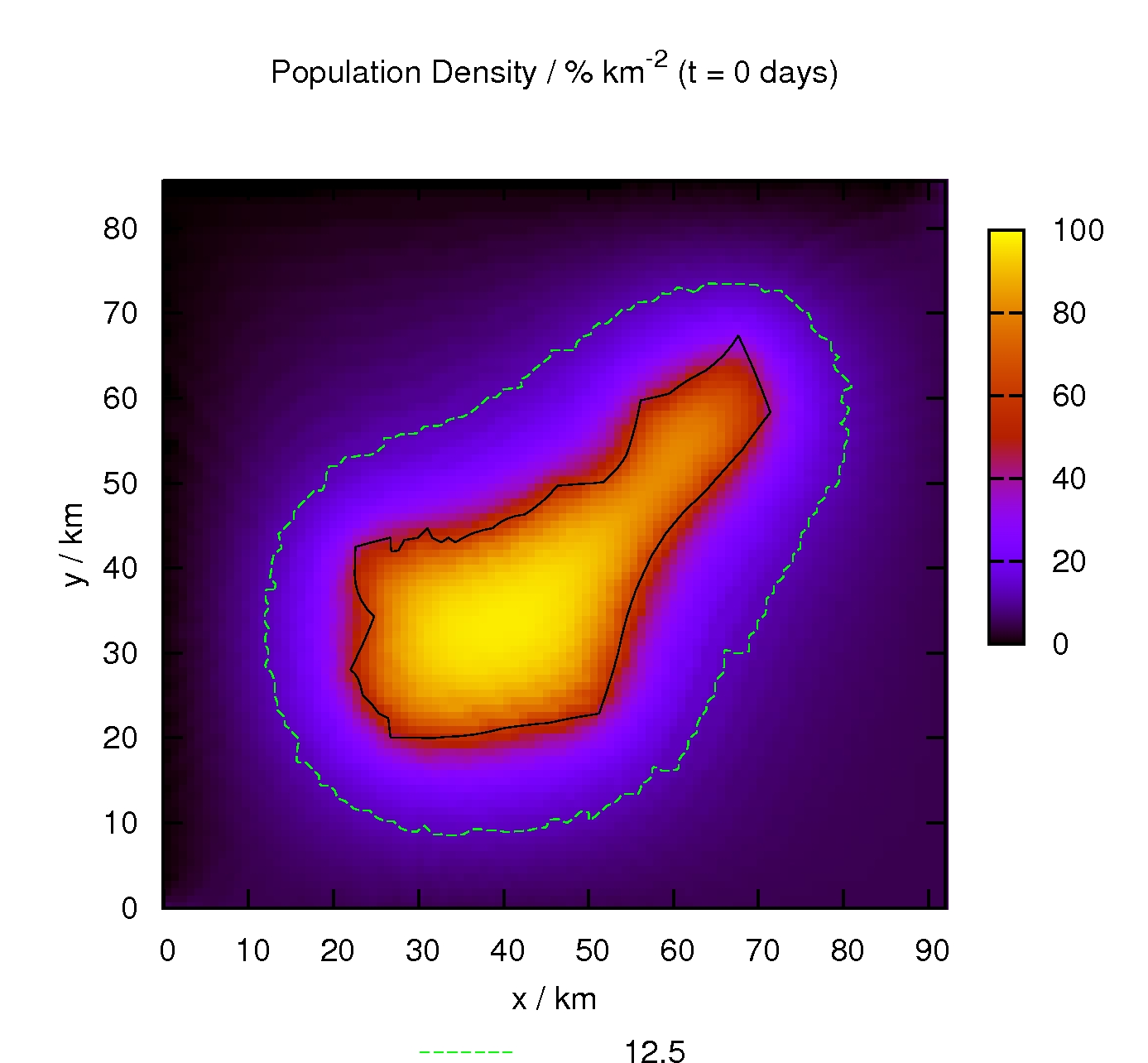}
\includegraphics[width=7.7cm, angle=0, clip = true]{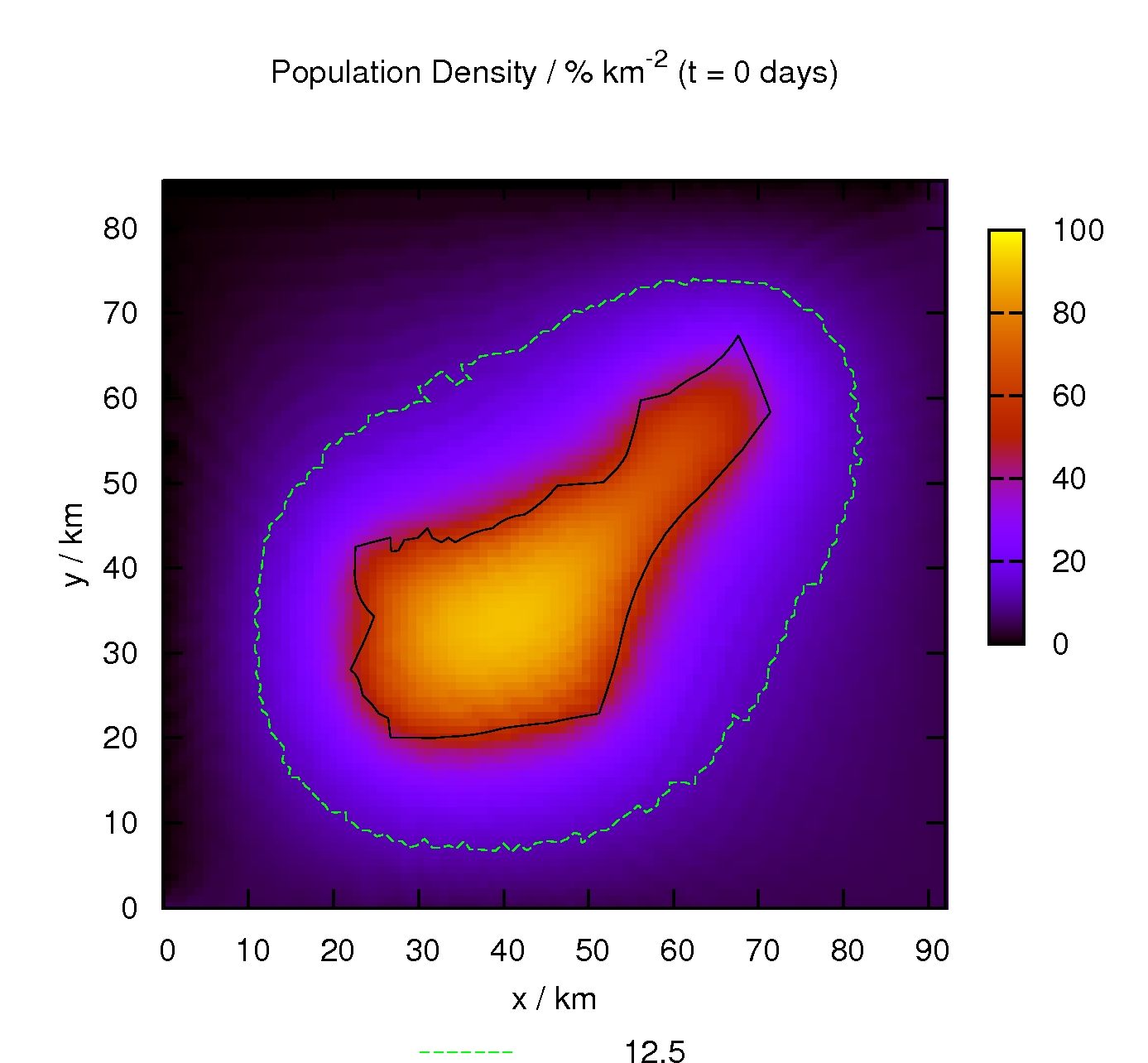}
\caption{Computed steady states for $\lambda = 0.16 \ \mathrm{km}^2 \
\mathrm{day}^{-1}$ (at left) and $\lambda = 0.32 \ \mathrm{km}^2 \
\mathrm{day}^{-1}$ (at right) using {\em G. brevipalpis} carrying capacities.
Notice that the boundary conditions are starting to effect the computed zone of
influence at these high diffusion rates.} \label{2nd}
   \end{center}
\end{figure}
No information, whatsoever, would appear to be available on the diffusion rates
of either {\em G. brevipalpis} or {\em G. austeni}. The most comprehensive set
of measurements are probably those for {\em G. morsitans}, recorded by {\sc
Jackson} \cite{Jackson1} (reported in {\sc Rogers} \cite{Rogers1}). The rate at
which {\em G. morsitans} dispersed in {\em G. swynnertoni} habitat was found to
be a consistant 0.153 $\mathrm{km}^2$ $\mathrm{day}^{-1}$, so long as the
initial stages of the experiment are omitted. Other work, mostly by the same
author (also summarised in {\sc Rogers} \cite{Rogers1}), suggests {\em G.
morsitans} was possibly uncomfortable in {\em G. swynnertoni} habitat. The very
low end of the {\em G. morsitans} range would appear to be about 0.04
$\mathrm{km}^2$ $\mathrm{day}^{-1}$. The point is that different habitats can
have different coefficients, as do different species, and one would imagine
temperature plays a role. {\em G. brevipalpis} and {\em G. austeni} are
profoundly different species to {\em G. morsitans}, in both size and habitat.
{\em G. morsitans} is of an intermediate size, while {\em G. brevipalpis} is one
of the largest flies known. {\em G. austeni} is the smallest of the tsetse
flies. {\em G. brevipalpis} and {\em G. austeni} are both forest-dwelling,
whereas {\em G. morsitans} is a savannah species.

The lack of known parameters need not necessarily be cause for despair.
Worst-case values are sought and such values are readily deduced from Figure
\ref{risk}. The premise of this work is that the reserve is a problem, that it
is the cause of unusually high tsetse numbers in the adjacent agricultural
areas. Indeed, the difference in habitat, visibly discernable in satellite
images, and Figure \ref{risk} are certainly suggestive of a haven with a zone of
influence, in the case of a very habitat-specific {\em G. brevipalpis}. The
distribution of {\em G. austeni} is less well understood as there appear to be
areas of good {\em G. austeni} habitat outside the reserve. It is arguable
whether an analogous, likely zone of influence can be detected about the
reserve; that is until one casts one's eyes on the St. Lucia region, to the
east, for corroboration. 

Worst-case diffusion coefficients for both {\em G. brevipalpis} and {\em G.
austeni} were revealed by a strategy of trial and error. Different values were
used until matching zones of influence to those evidenced by Figure \ref{risk}
were produced. The value of the diffusion coefficient was either halved, or
doubled, until a suitable zone of influence was generated. The transition
through 12.5 \% was deemed the most distinct in the {\em G. brevipalpis} data.
The final matches, Figures \ref{1st} and \ref{2nd}, were obviously not perfect
due to higher ground to the north, suitable habitat outside the reserve (which
was not modelled) and supplementation from the St. Lucia side of the reserve.
With hindsight, a bigger domain would have been preferred. The crude technique,
nonetheless, suggested some very acceptable values. 

\renewcommand{\thefootnote}{\fnsymbol{footnote}}
The diffusion coefficient for {\em G. austeni} is probably 0.04 $\mathrm{km}^2$
$\mathrm{day}^{-1}$, that of a very comfortable {\em G. morsitans} population.
Intuitively, this makes sense because of the small size of the species. At
worst, one might speculate that it could reach 0.08 $\mathrm{km}^2$
$\mathrm{day}^{-1}$. The diffusion coefficient for {\em G. brevipalpis},
however, came as something of a surprise for a forest-dwelling species. At
somewhere between 0.16 $\mathrm{km}^2$ $\mathrm{day}^{-1}$ and, very possibly as
high as, 0.32 $\mathrm{km}^2$ $\mathrm{day}^{-1}$, it approaches {\sc Rogers}
\cite{Rogers1}'s observations of {\em G. fuscipes fuscipes}, a fly of similar
habitat, though smaller size.
\renewcommand{\thefootnote}{\arabic{footnote}}

\subsection{The Temperature, $T$}

The South African Meteorological Services quote a mean annual temperature of
22.1 $^\circ$C for Mpila, inside the reserve (based on data collected during the
1980's and 1990's). This is consistant with the data of {\sc Schulze} and {\sc
Maharaj} \cite{SchulzeAndMaharaj}, who define the overall area as being of a
temperature greater than 22 $^\circ$C. The Mpila value is further corroborated
by ARC-ISCW automatic weather stations situated between Mtunzini and Pongola
(operational since 2004). They suggest an average daily temperature of 22
$^\circ$C, according to the Department of Agriculture. The daily temperature for
the entire region was taken to be the mean annual temperature of Mpila, 22.1
$^\circ$C. 

\subsection{The Finite Element Mesh}

Nine-noded quadrilateral elements and the associated $Q_2$ element basis were used. A program to generate the mesh was written based on {\sc Childs} and {\sc Reddy} \cite{Childs3}.

\begin{figure}[H]
\vspace{-10mm}
    \begin{center}
\includegraphics[height=15cm, angle=0, clip = true]{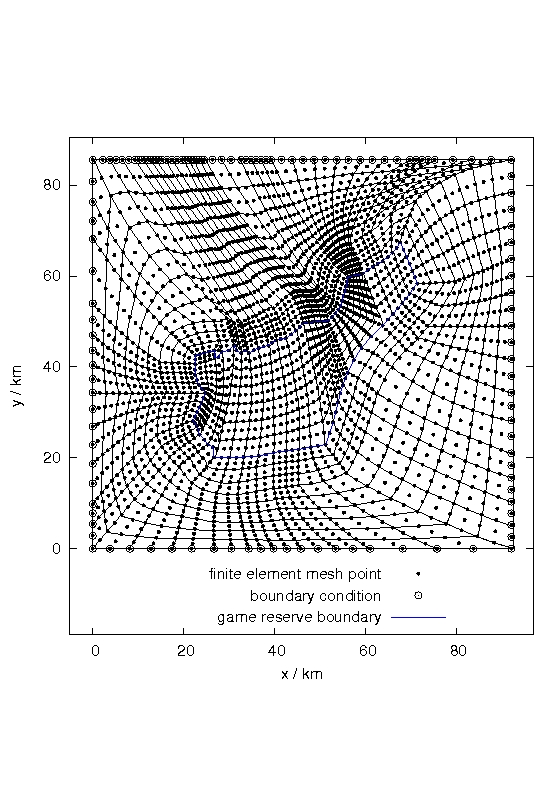}
\caption{The finite element mesh.} \label{mesh}
   \end{center}
\vspace{-10mm}
\end{figure}

\subsection{Boundary Conditions}

The population was assumed to vanish along the northern and western boundaries.
A constant supply of flies, maintaining the eastern and southern boundaries of
the entire region at a 5\% level, was assumed. 

\section{Simulation}

The topic of simulation has already been broached in determining diffusion coefficients. The simulations attempted to address the following questions, which were identified as relevant to the Hluhluwe-iMfolozi problem:
\begin{enumerate}
\item What are the likely, worst-case diffusion coefficients for {\em G. brevipalpis} and {\em G. austeni}? (Completed in Section \ref{diffusion})
\item What will the long-term effect of the temporary, 2-year use of pour-ons in the surrounding areas be?
\item Can the populations within the reserve be `siphoned out' to extinction from outside the reserve; failing that, down to the 20\% level? 
\item What is a practical barrier width?
\item Can the influence of the Hluhluwe-iMfolozi Game Reserve on surrounding
tsetse population levels be neutralized? 
\item Can the tsetse populations of the Hluhluwe-iMfolozi Game Reserve and
their associated trypanosomiasis be isolated from the surrounding areas? 
\item In what way are diffusion rates relevant to containment, eradication and any subsequent rebound?
\end{enumerate}

The initial, start-up values were taken to be the carrying capacities, with the
exception of the northern and western boundaries, where the population was
assumed to vanish. For each simulation the model was first run for two years;
more than adequate time to allow it to settle down to the steady-state from its
start-up values. The model was then run for another two
years\footnotemark[1]\footnotetext[1]{Deltamethrin pour-ons can not safely be
used on cattle for any longer than two years without compromising their
resistance to tick-bourne diseases, consequently an enzootic condition.} with
various controls in place. In instances in which the controls were either
revised or removed, the population was allowed to rebound for a further one or
two years. Unless otherwise stated, {\em G. austeni} carrying capacities were
used in association with diffusion rates of 0.08 $\mathrm{km}^2$
$\mathrm{day}^{-1}$ and below, while {\em G. brevipalpis} carrying capacities
were used in association with diffusion rates of 0.16 $\mathrm{km}^2$
$\mathrm{day}^{-1}$ and above. 

Barriers of a width greater than 5 $\mathrm{km}$  were not experimented with,
even though they can be expected to be more optimal in terms of the required
number of targets. This is since they were deemed to be a self-defeating waste
of land. The barriers were modelled in such a way that their quoted widths
usually included a reasonable safety margin. This built-in, safety margin was
increased substantially ($\approx$ -15\%) along the northern, concave boundary
of the reserve, for the purposes of a crude sensitivity analysis. Incorrect
barrier construction, theft, storms, malfunction and fires (such as are visible
in Figure \ref{HluhluweImfoloziSatelliteImage}) are all eventualities which must
be prepared for. A reduction of the Figure \ref{1st} and \ref{2nd} tsetse levels
by an order of magnitude was adopted as a guideline for the tolerance used in
evaluating barriers. (It is reasoned that if the incidence of nagana due to
flies from the game reserve was 10 \% for a given time frame, a barrier with
such a tolerance should cause it to drop below 1 \%.) Thus, a zone with a
population density lower than \mbox{4 \% $\mathrm{km}^{-2}$} is considered to be
substantially vacant. No flies are able to leave a vacant zone, which, in turn,
means no flies ever cross one of the size in question. The value of \mbox{4 \%
$\mathrm{km}^{-2}$} is also in keeping with {\sc Hendrickx} \cite{Hendrickx}'s
lowest detectable level, the \mbox{0\%} to \mbox{6.3\%} category. 
\newpage

\subsection{Results}

\subsubsection*{A 2.5 $\mathrm{km}$-Wide Versus a 5 $\mathrm{km}$-Wide Barrier Surrounding the Reserve}
 
\begin{figure}[H]
    \begin{center}
\includegraphics[width=7.7cm, angle=0, clip = true]{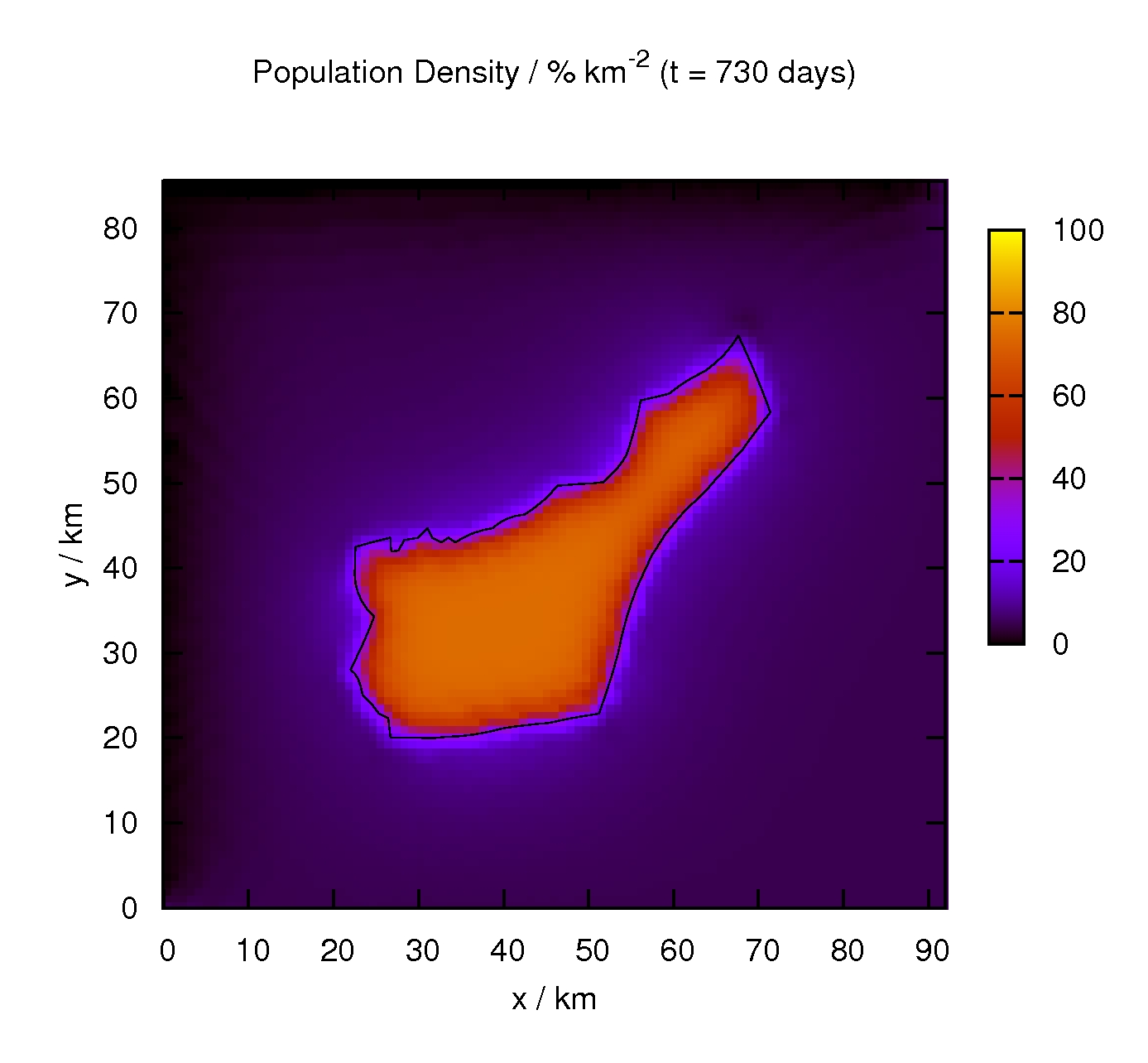}
\includegraphics[width=7.7cm, angle=0, clip = true]{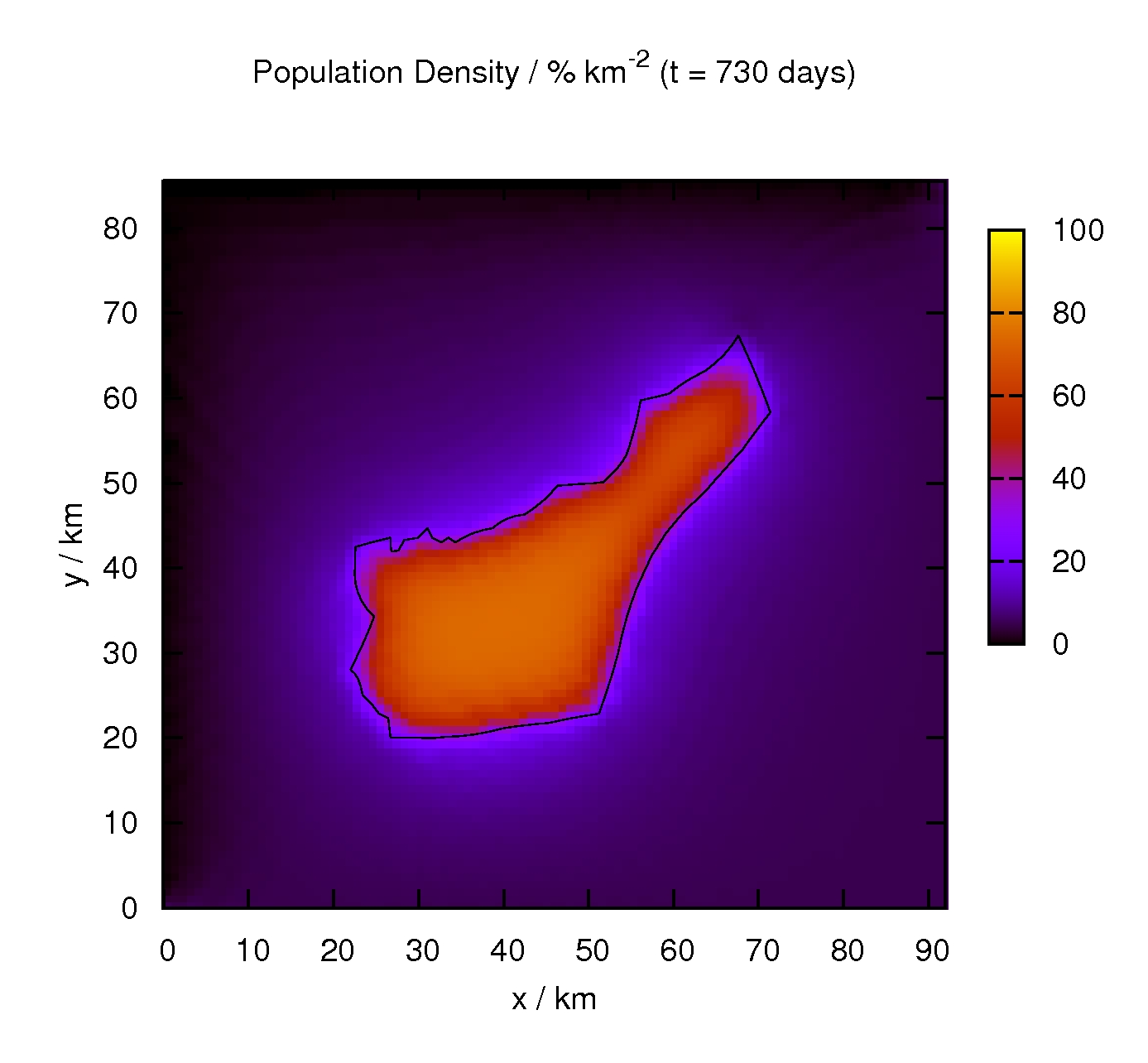}
\caption{The effect of an approximately 2.5 $\mathrm{km}$-wide barrier, with a
\mbox{2 \% $\mathrm{day}^{-1}$} mortality throughout, after 2 years. At left,
$\lambda = 0.04 \ \mathrm{km}^2 \ \mathrm{day}^{-1}$. At right, $\lambda = 0.08
\ \mathrm{km}^2 \ \mathrm{day}^{-1}$. A \mbox{2.5 $\mathrm{km}$-wide} barrier
is clearly not efficacious.} \label{2.5km}
\vspace{5mm}
\includegraphics[width=7.7cm, angle=0, clip = true]{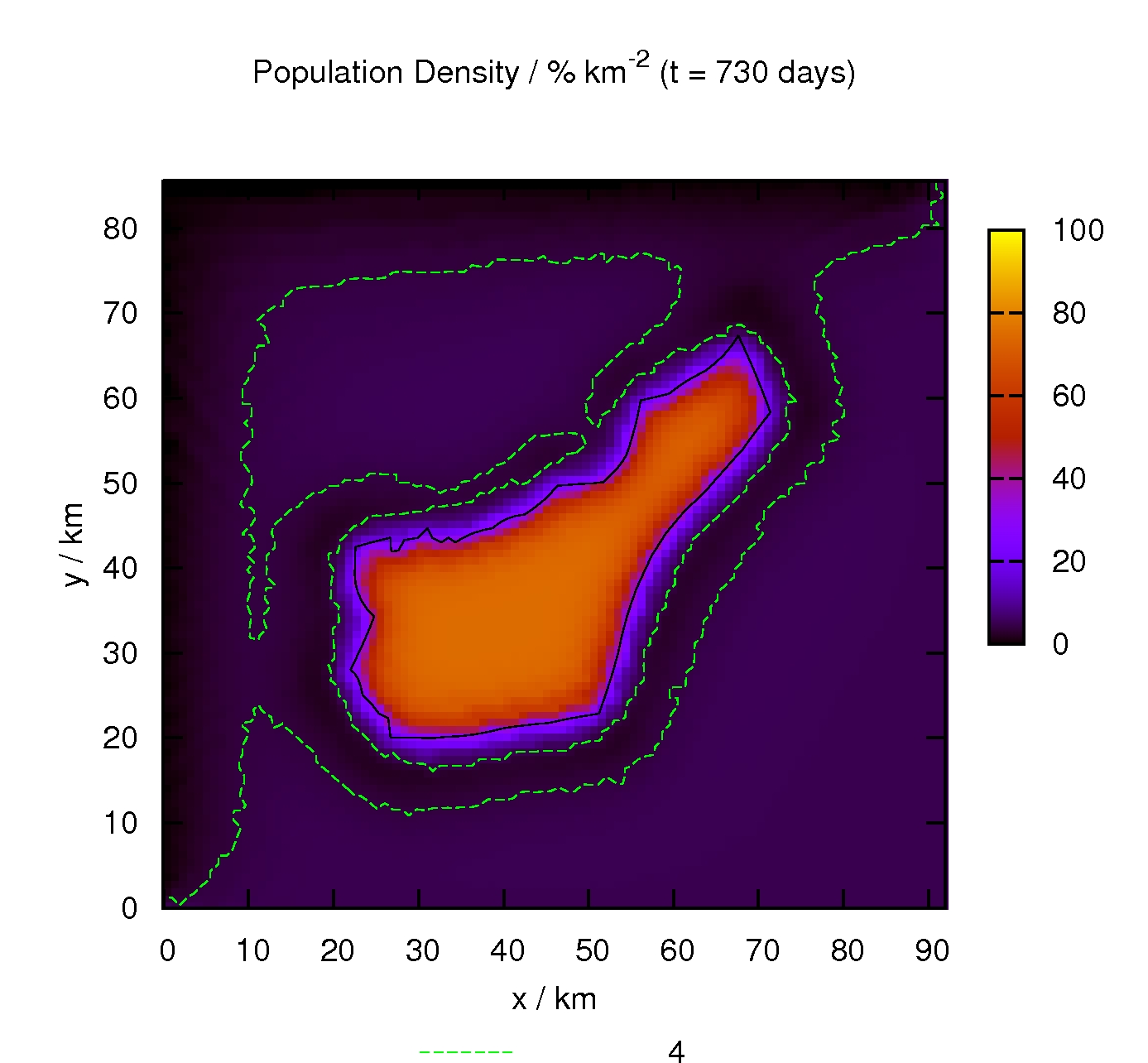}
\includegraphics[width=7.7cm, angle=0, clip = true]{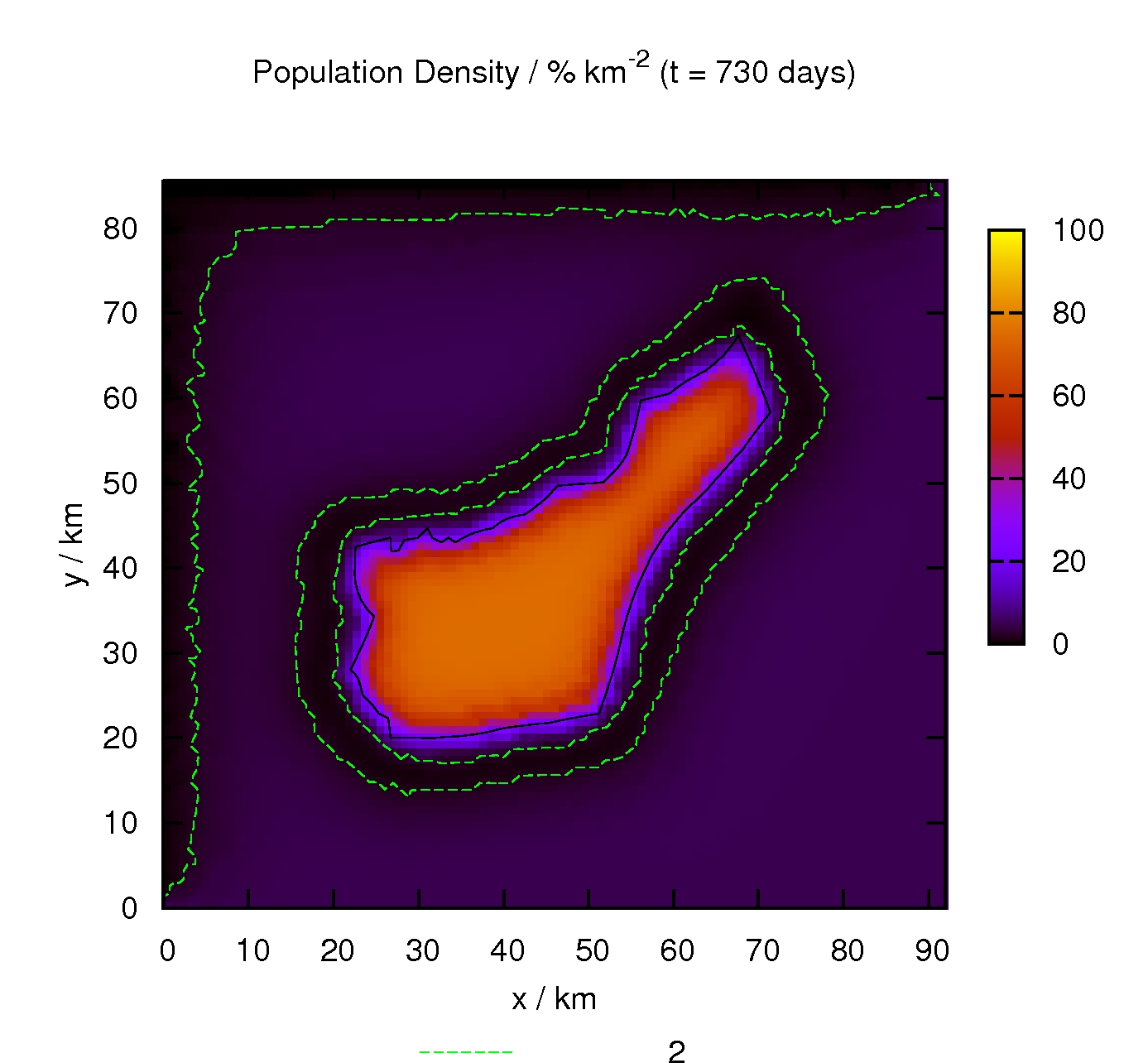}
\caption{The  effect of an approximately 5 $\mathrm{km}$-wide barrier, after 2
years, using \mbox{$\lambda = 0.04 \ \mathrm{km}^2 \ \mathrm{day}^{-1}$}. At
left, a barrier with a mortality of \mbox{2 \% $\mathrm{day}^{-1}$} throughout.
At right, a barrier with a mortality of \mbox{4 \% $\mathrm{day}^{-1}$}
throughout. The 5 $\mathrm{km}$-wide barrier has already successfully isolated
the {\em G. austeni}, reserve population, should it have a diffusion rate as low
as 0.04 $\mathrm{km}^2$ $\mathrm{day}^{-1}$.} \label{5kmA}
   \end{center}
\end{figure}

\subsubsection*{The Quest for an Impenetrable, Surrounding Barrier}

\begin{figure}[H]
    \begin{center}
\includegraphics[width=7.7cm, angle=0, clip = true]{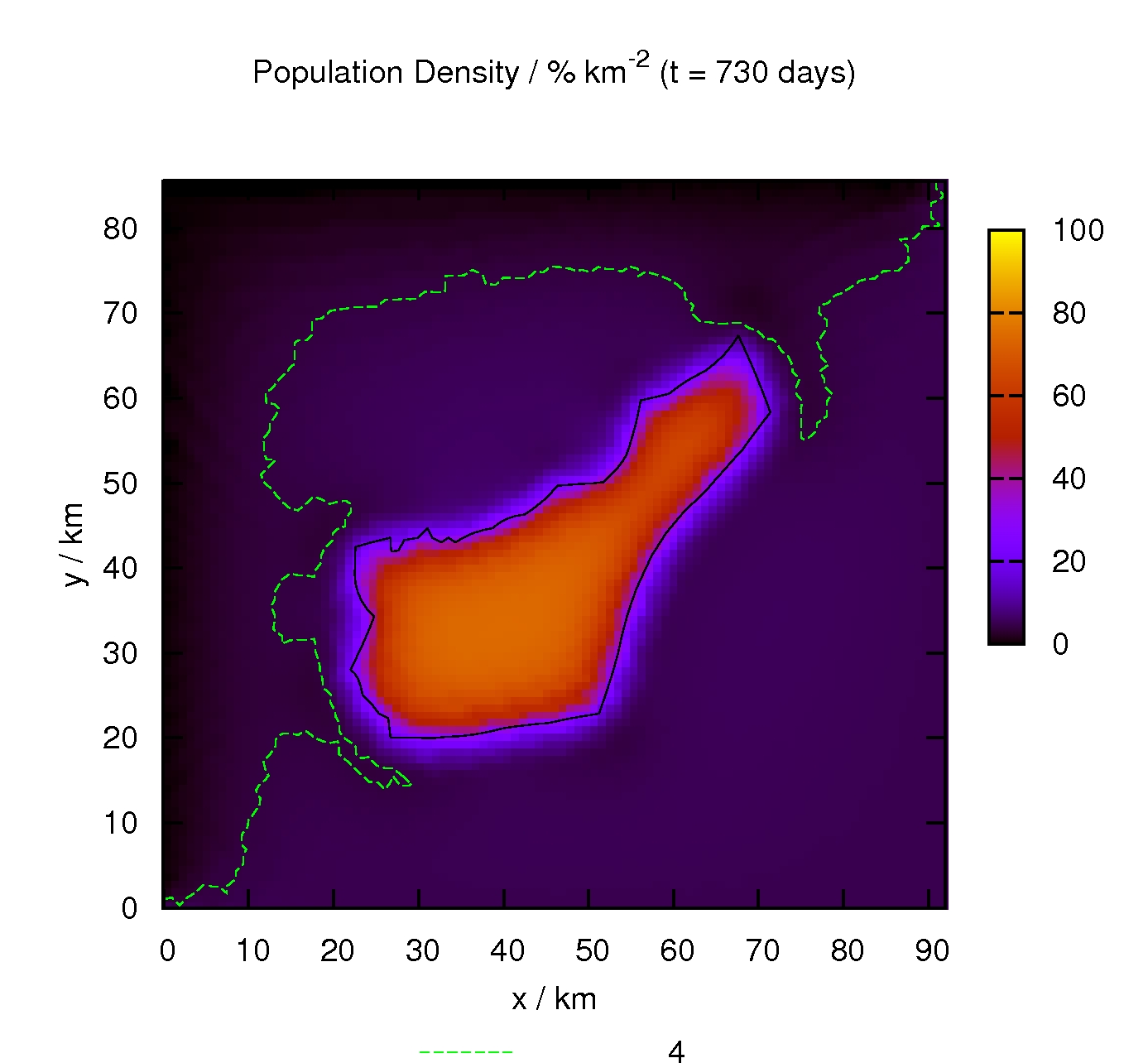}
\includegraphics[width=7.7cm, angle=0, clip = true]{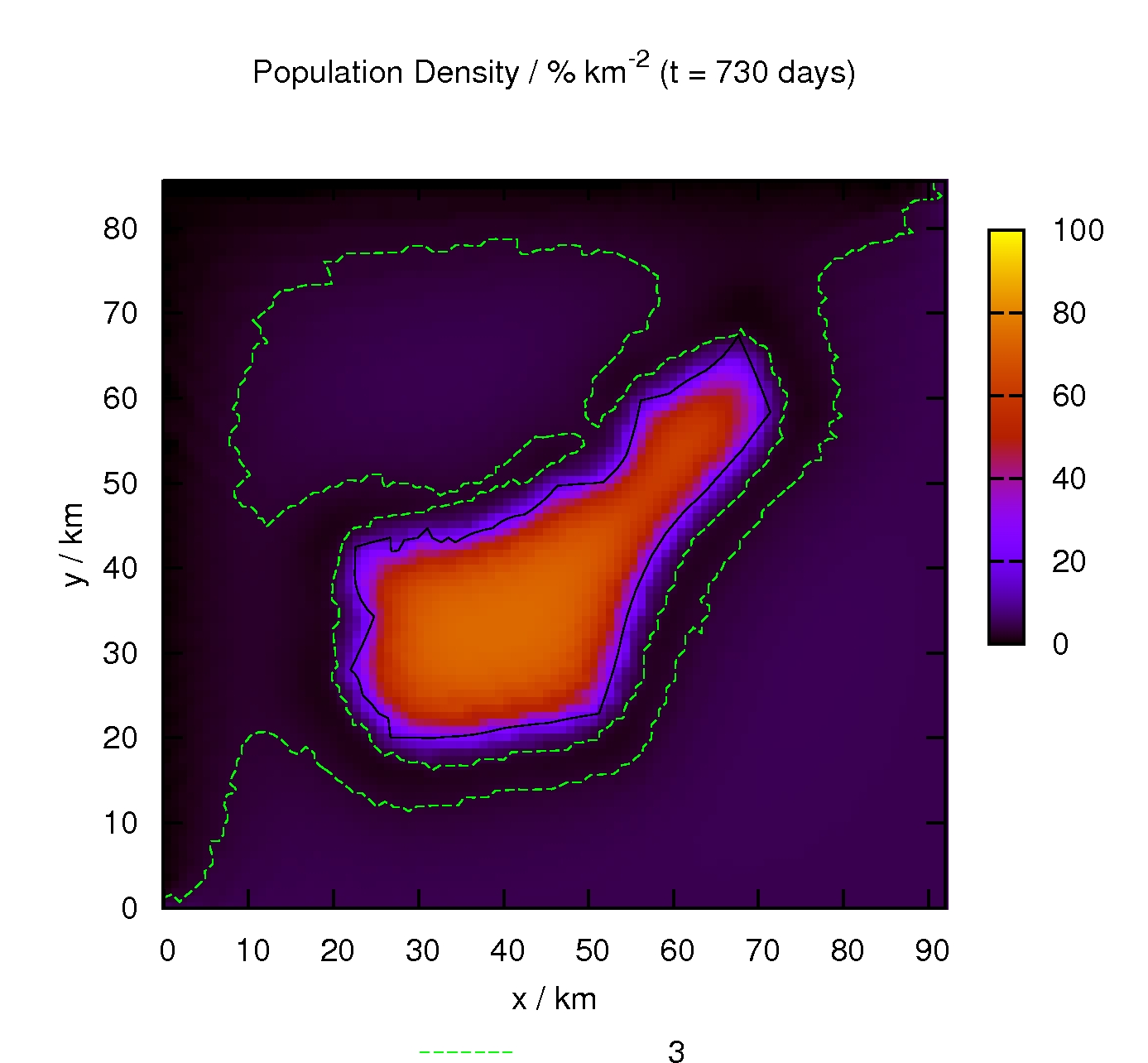}
\caption{The  effect of an approximately 5 $\mathrm{km}$-wide barrier, after 2
years, using \mbox{$\lambda = 0.08 \ \mathrm{km}^2 \ \mathrm{day}^{-1}$}. At
left, a 2 \% $\mathrm{day}^{-1}$ mortality throughout the barrier neutralizes
the influence of the reserve on surrounding tsetse populations, however, flies
with a greater prevalence and more lethal strains of trypanosomiasis are still
able to commute. At right, using a \mbox{4 \% $\mathrm{day}^{-1}$} mortality
throughout, the barrier has isolated the reserve.} \label{impenetrableA}
\vspace{5mm}
\includegraphics[width=7.7cm, angle=0, clip = true]{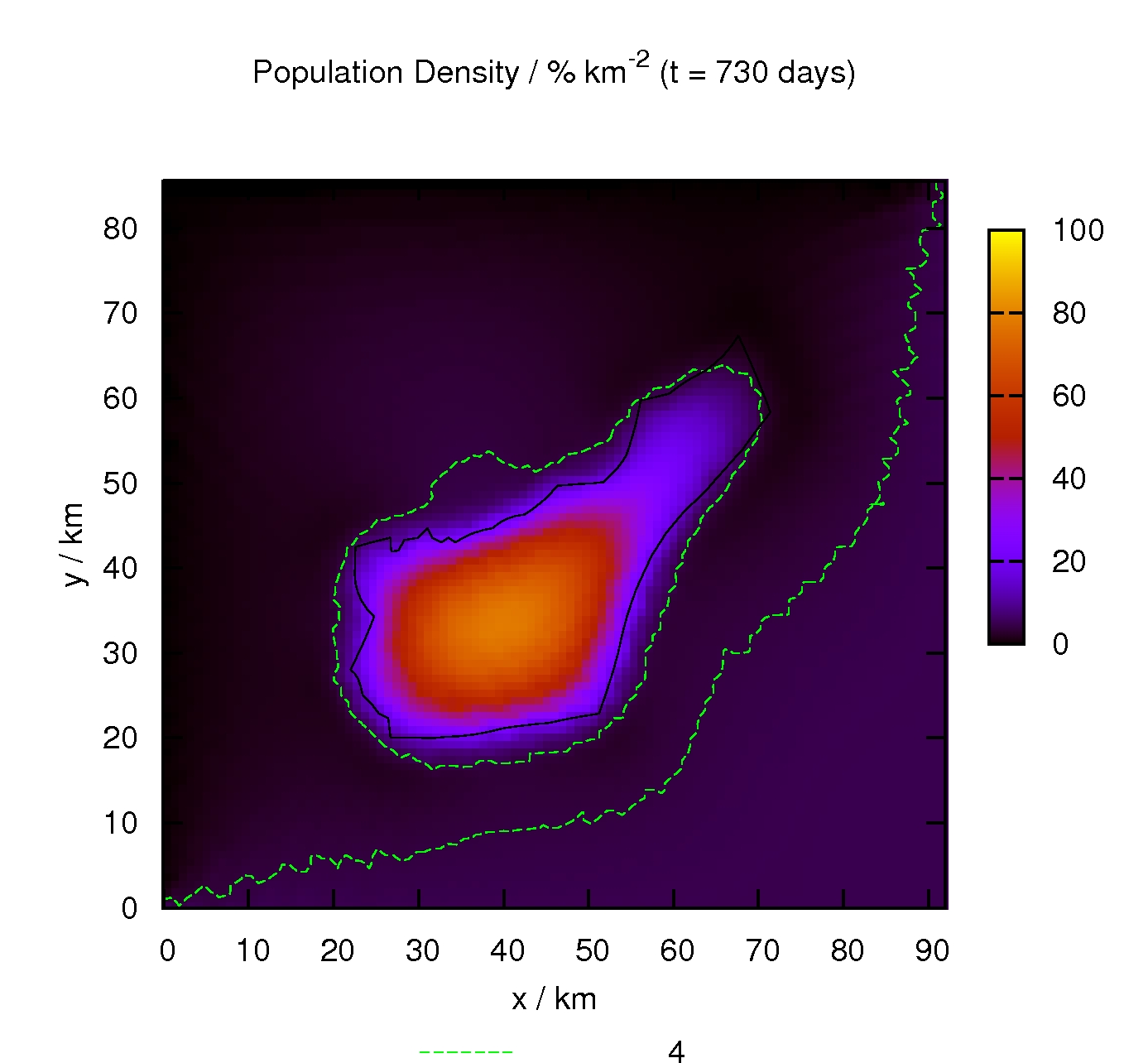}
\includegraphics[width=7.7cm, angle=0, clip = true]{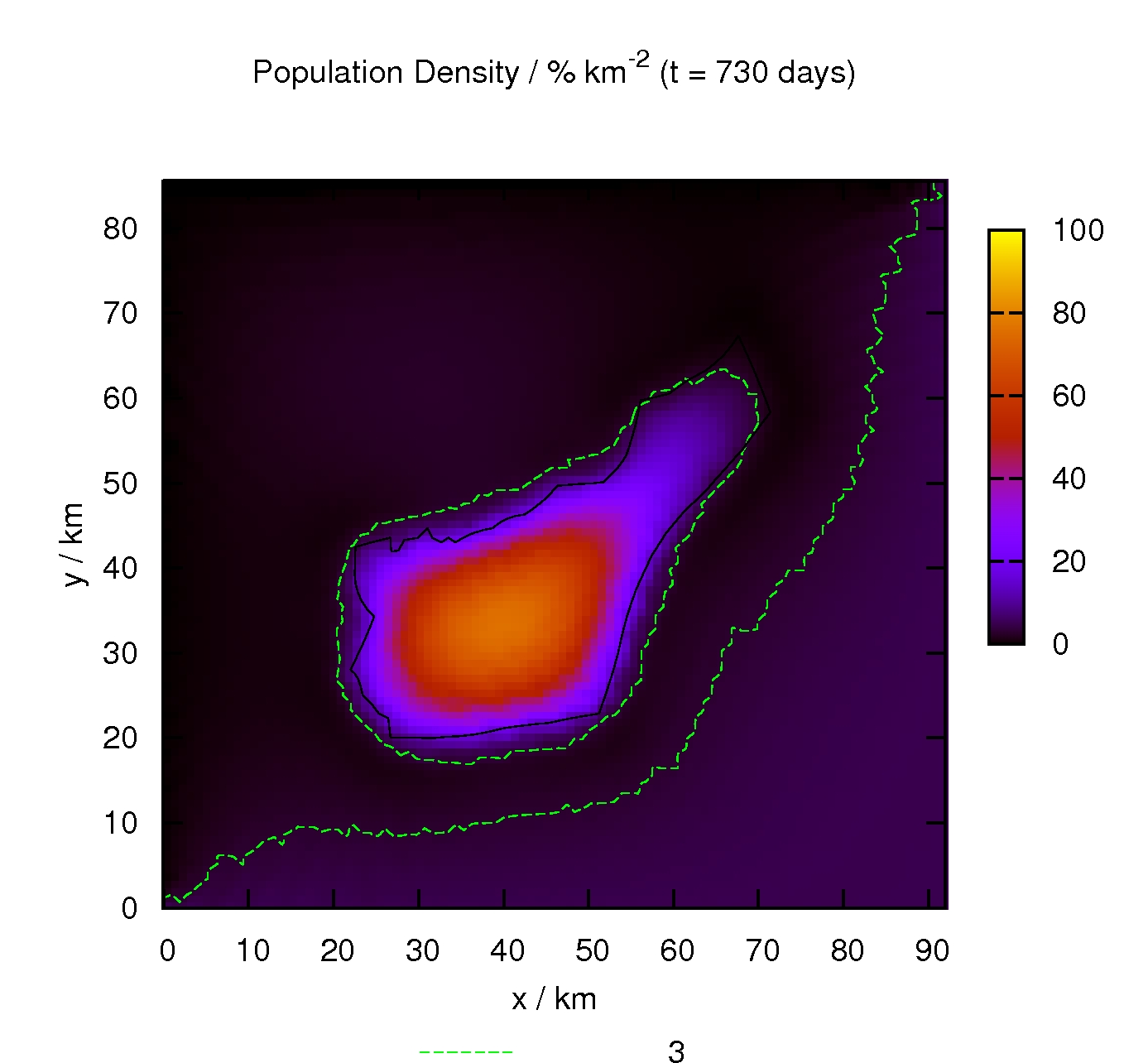}
\caption{The  effect of an approximately 5 $\mathrm{km}$-wide barrier, after 2
years, using \mbox{$\lambda = 0.32 \ \mathrm{km}^2 \ \mathrm{day}^{-1}$}. At
left, an 8 \% $\mathrm{day}^{-1}$ mortality throughout the barrier just fails to
isolate the reserve and then only along that part of the barrier-zone
constructed to be 15\% narrower for the purposes of sensitivity analysis. At
right, using a 12 \% $\mathrm{day}^{-1}$ mortality throughout, the barrier has
isolated the reserve.} \label{impenetrableB}
   \end{center}
\end{figure}

\subsubsection*{Pour-Ons and the Subsequent Rebound}

\begin{figure}[H]
    \begin{center}
\includegraphics[width=7.7cm, angle=0, clip = true]{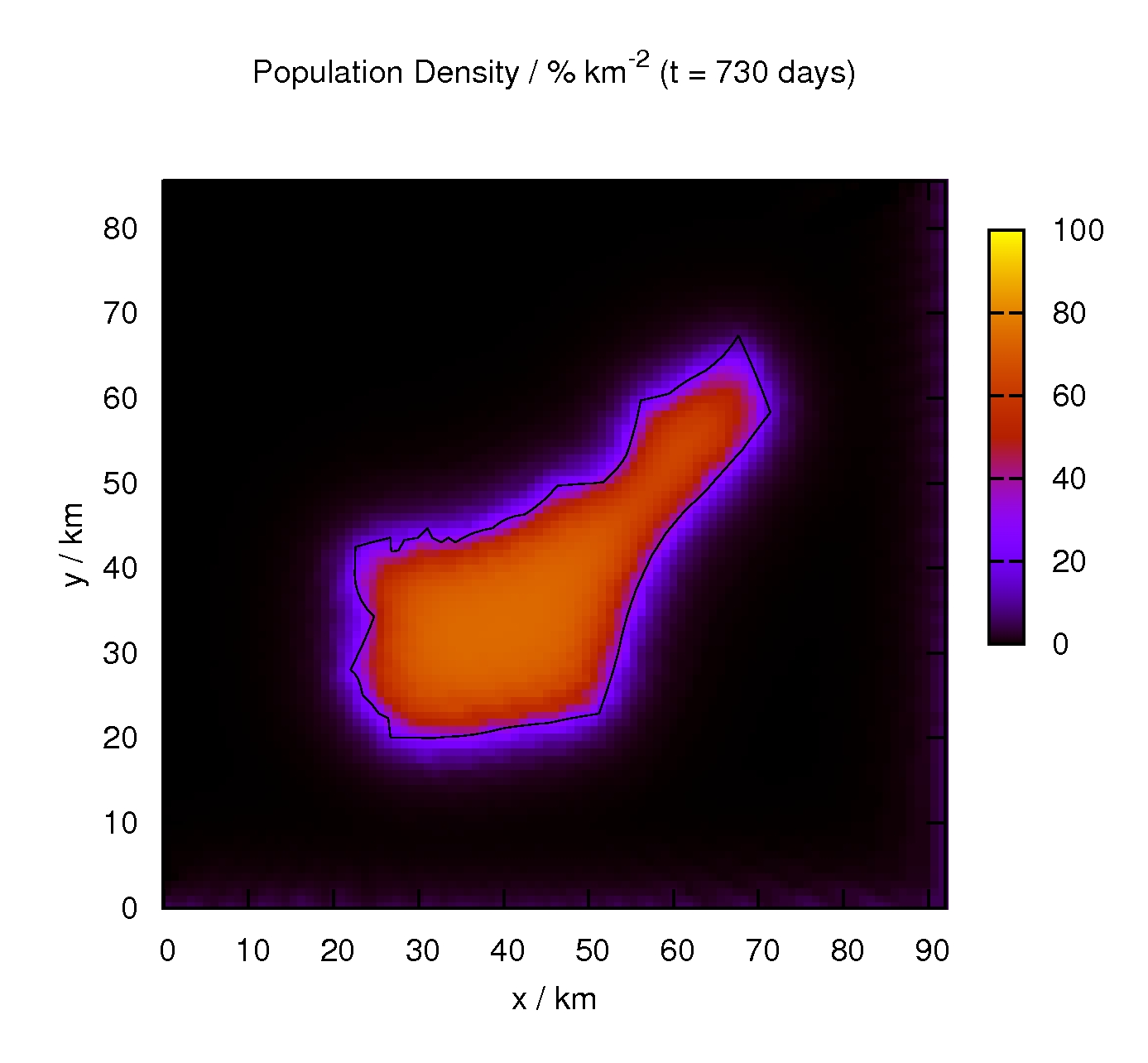}
\includegraphics[width=7.7cm, angle=0, clip = true]{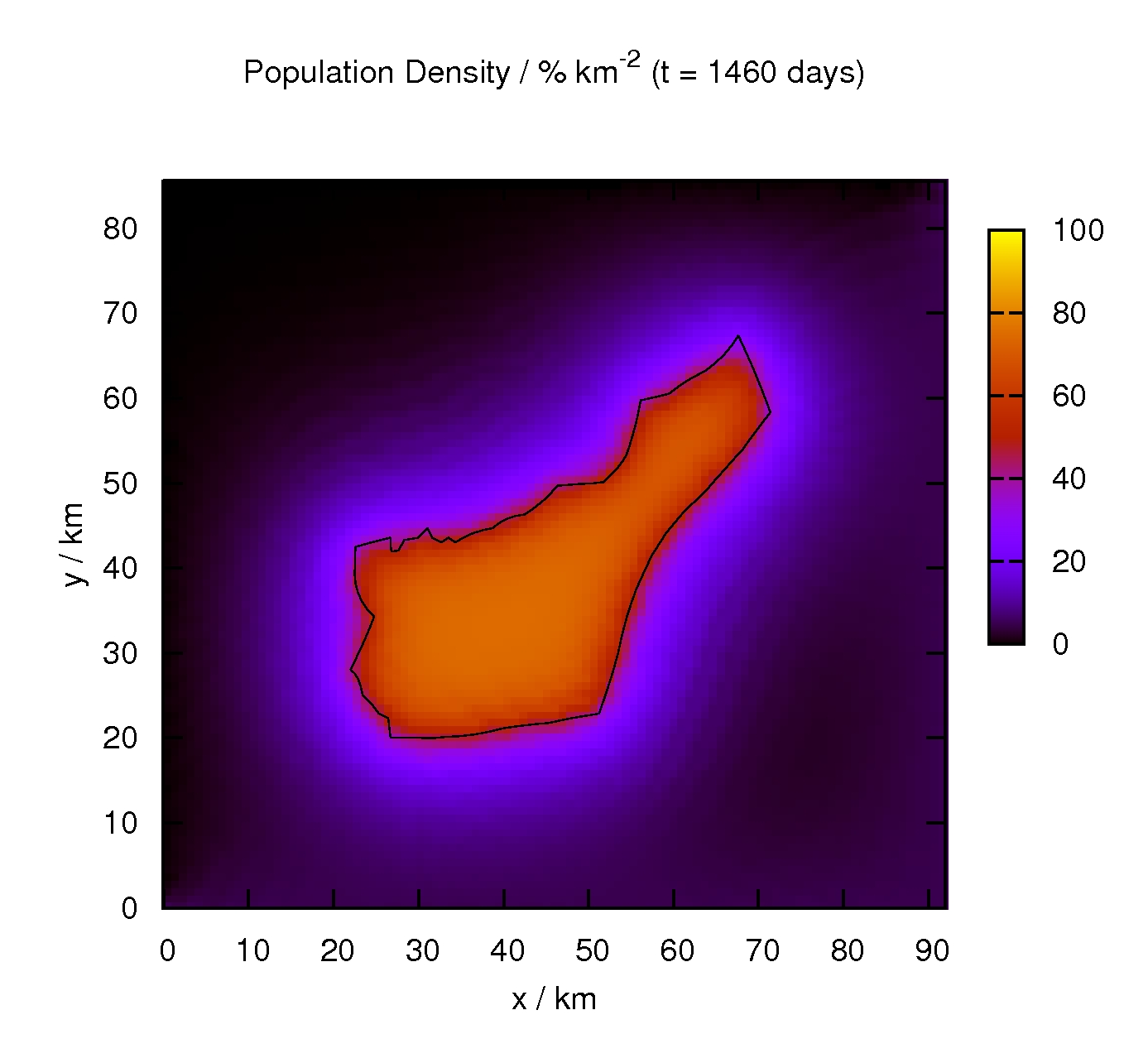}
\caption{The effect of a 2 \% $\mathrm{day}^{-1}$ mortality imposed everywhere
outside the reserve, for a period of 2 years, using \mbox{$\lambda = 0.08 \
\mathrm{km}^2 \ \mathrm{day}^{-1}$} (left); the rebound after a further 2 years
(right). There is a complete rebound into the areas immediately surrounding the
reserve, however, more remote areas of poor habitat do not recover as quickly.}
\label{pourOnsA1}
\vspace{5mm}
\includegraphics[width=7.7cm, angle=0, clip = true]{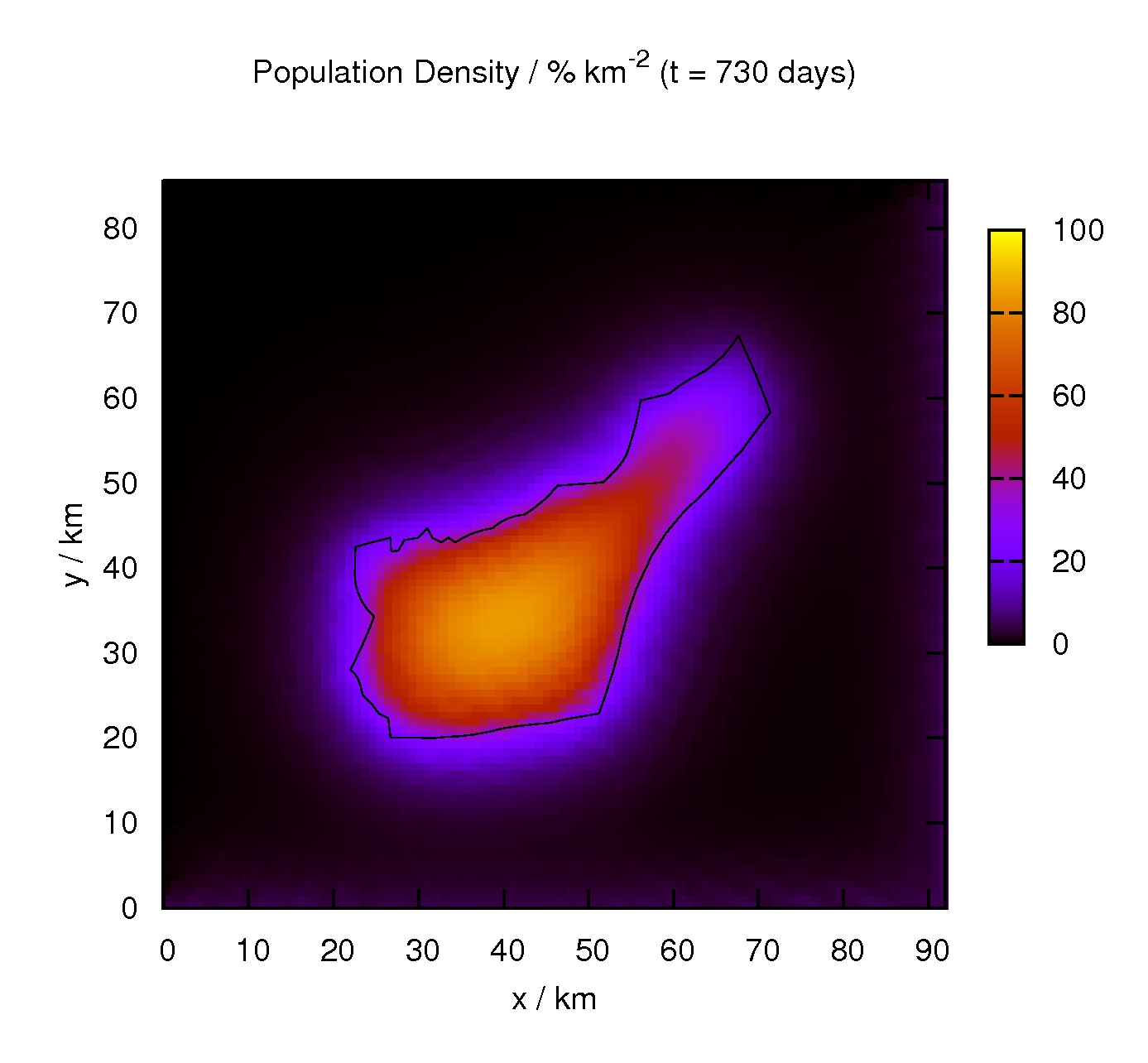}
\includegraphics[width=7.7cm, angle=0, clip = true]{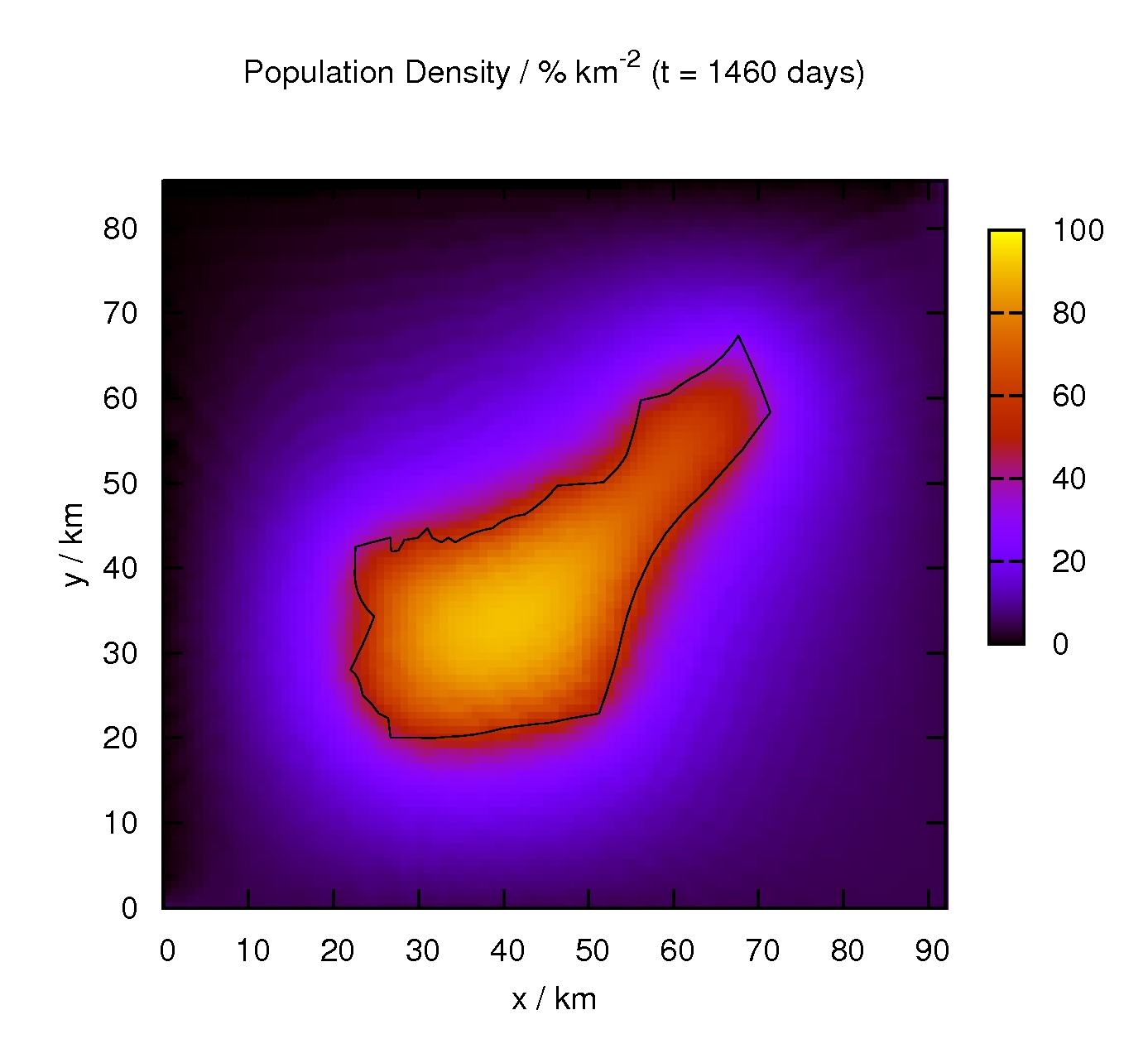}
\caption{The effect of a 2 \% $\mathrm{day}^{-1}$ mortality imposed everywhere
outside the reserve, for a period of 2 years, using \mbox{$\lambda = 0.32 \
\mathrm{km}^2 \ \mathrm{day}^{-1}$} (left); the rebound after a further 2 years
(right). There is a complete rebound.} \label{pourOnsB1}
   \end{center}
\end{figure}

\subsubsection*{The Implications of the Diffusion Coefficient for Eradication and Any Subsequent Rebound}

\begin{figure}[H]
    \begin{center}
\includegraphics[width=7.7cm, angle=0, clip = true]{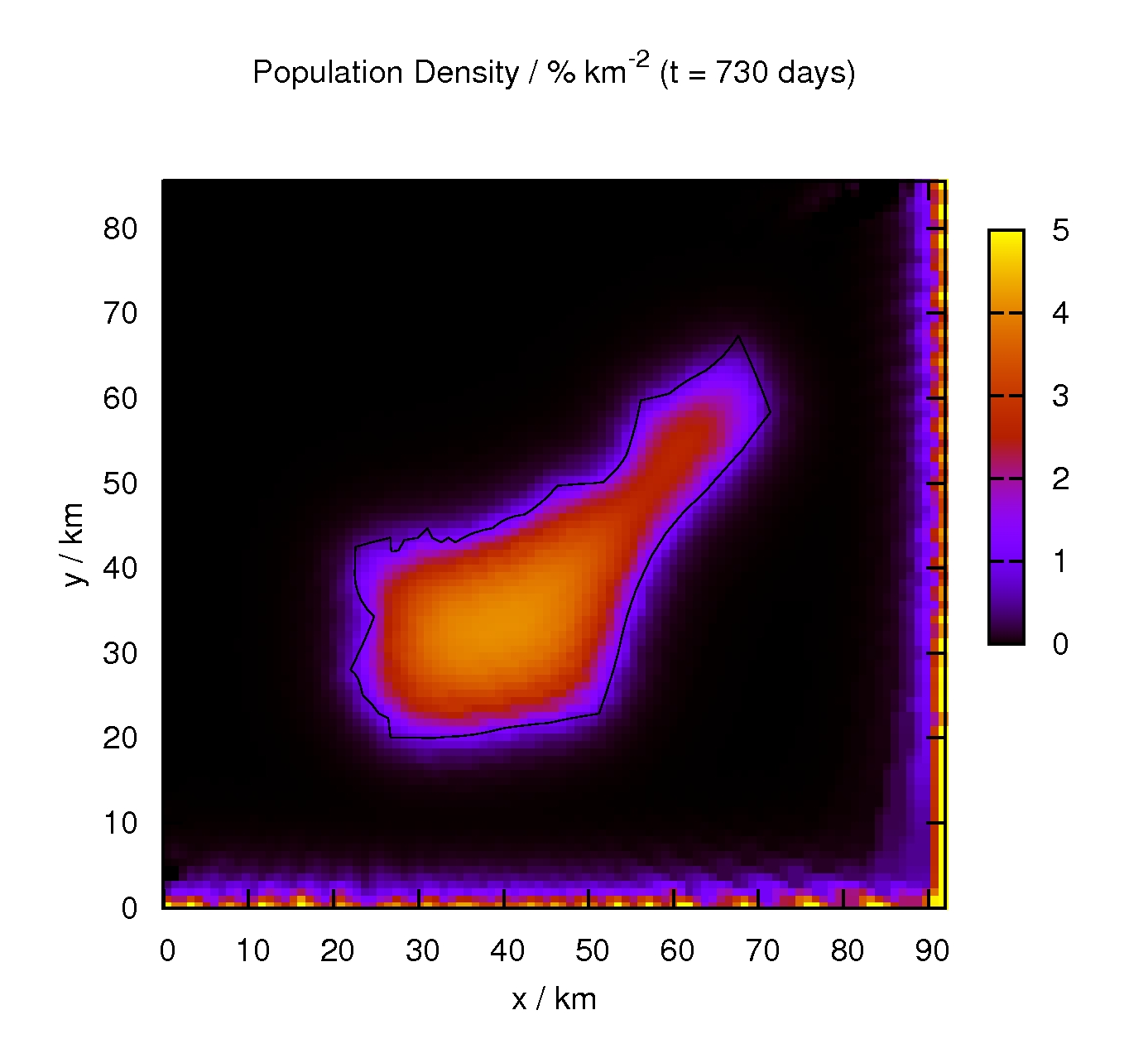}
\includegraphics[width=7.7cm, angle=0, clip = true]{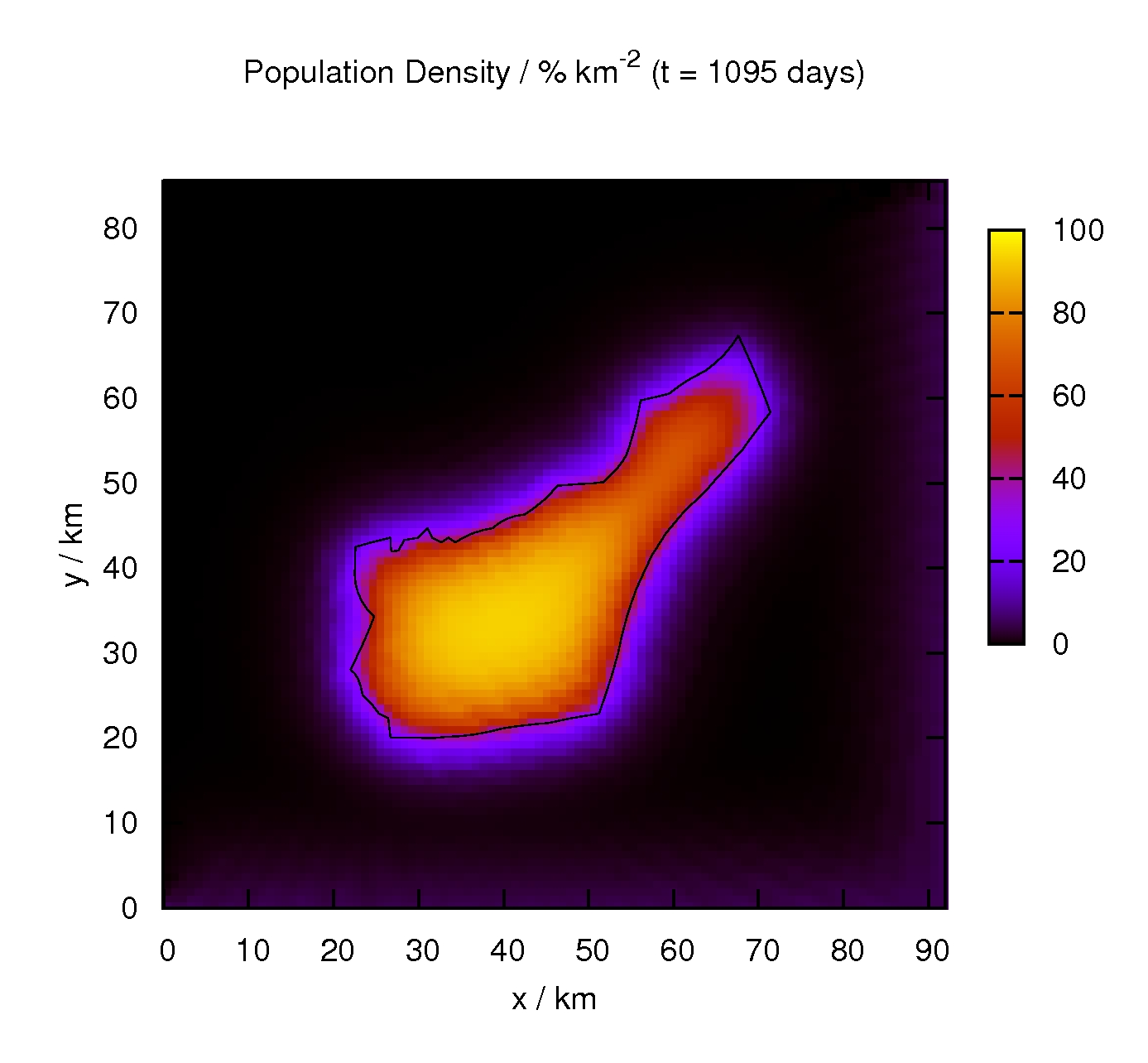}
\caption{The effect of a 2 \% $\mathrm{day}^{-1}$ mortality imposed everywhere
for a period of 2 years, using $\lambda = 0.04 \ \mathrm{km}^2 \
\mathrm{day}^{-1}$ (left); the population rebound after a further year (right).
{\em G. brevipalpis} carrying capacities were used for comparative purposes.}
\label{eradicationA}
\vspace{5mm}
\includegraphics[width=7.7cm, angle=0, clip = true]{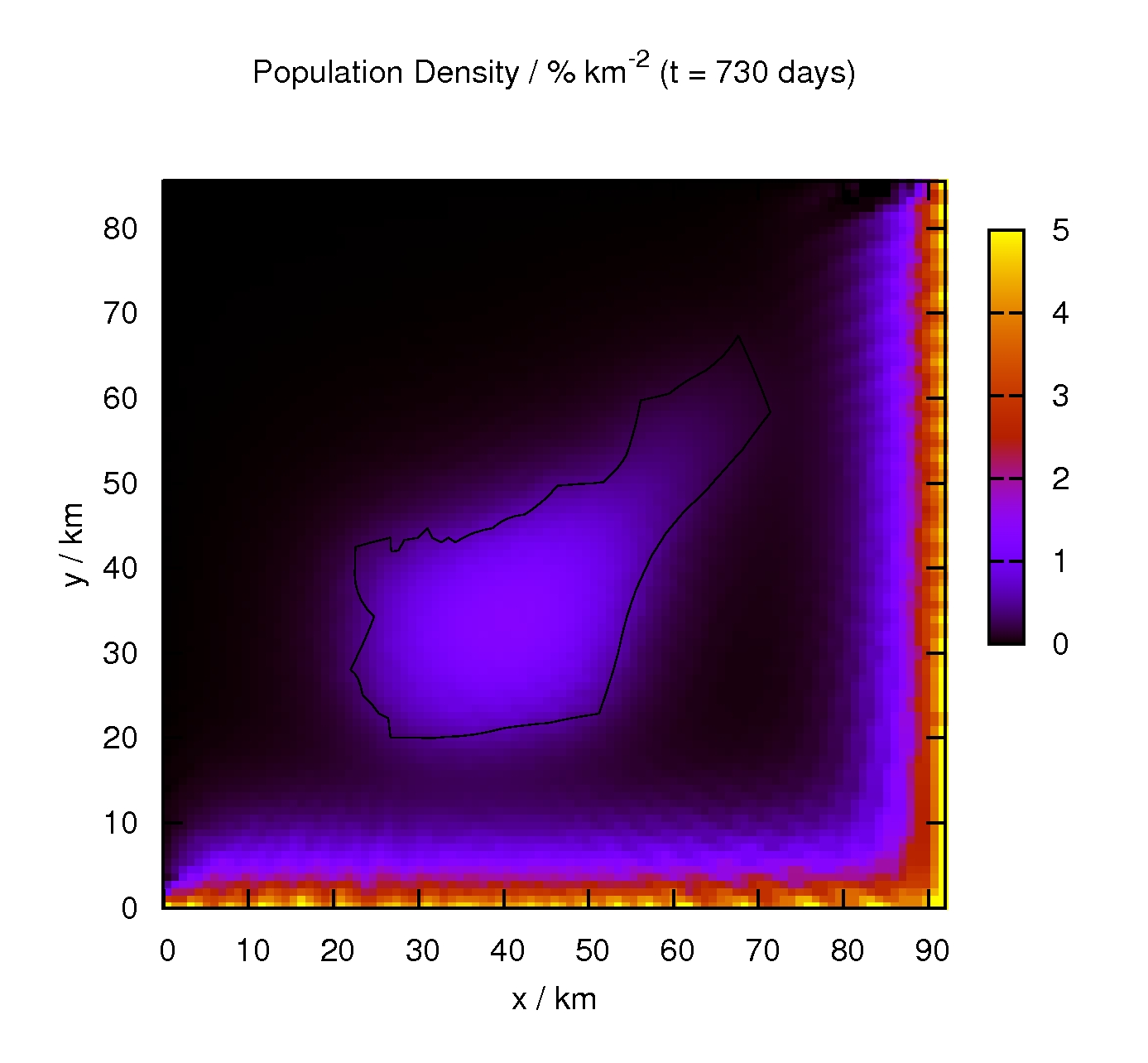}
\includegraphics[width=7.7cm, angle=0, clip = true]{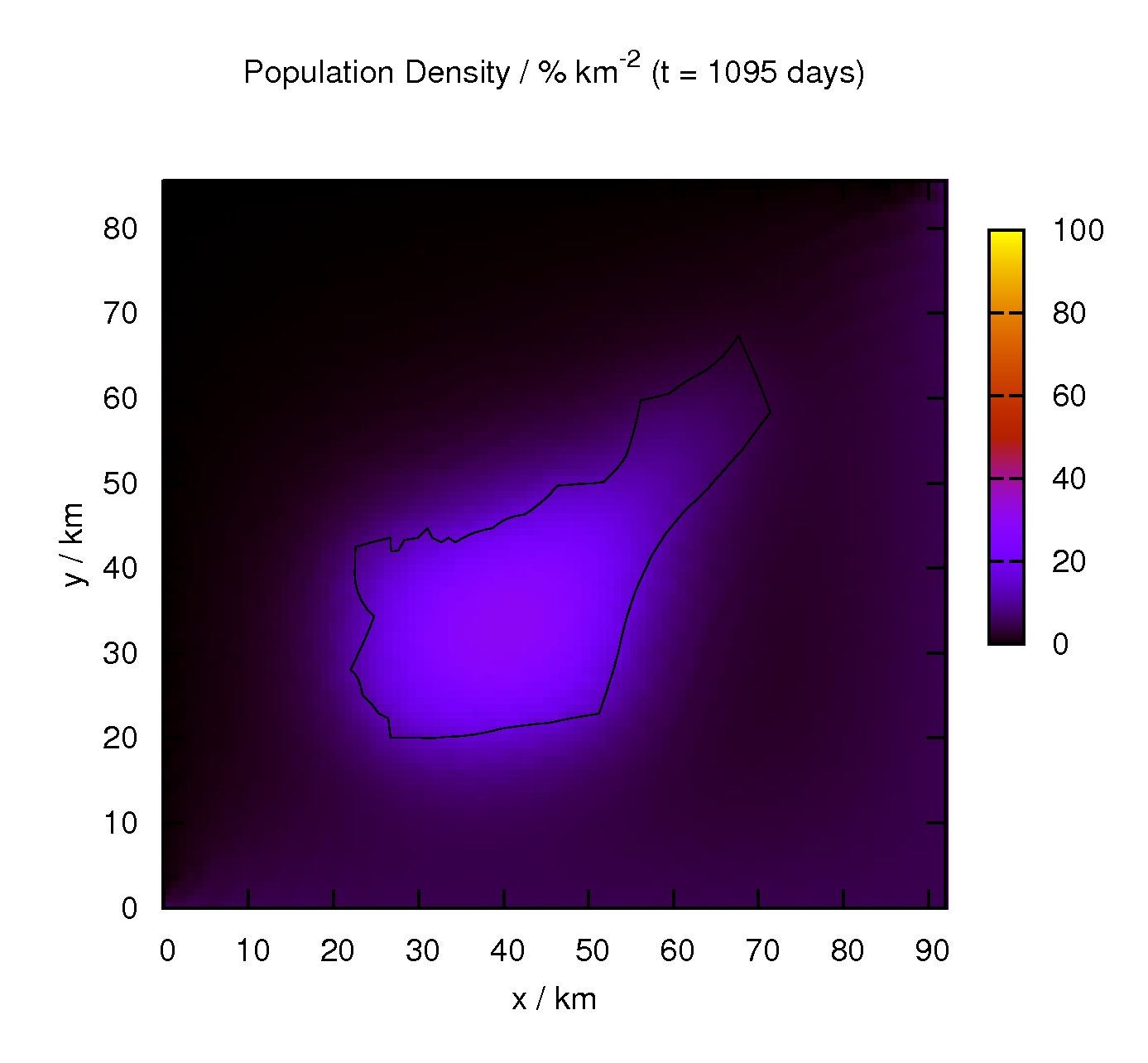}
\caption{The effect of a 2 \% $\mathrm{day}^{-1}$ mortality imposed everywhere
for a period of 2 years, using \mbox{$\lambda = 0.32 \ \mathrm{km}^2 \
\mathrm{day}^{-1}$} (left); the population rebound after a further year (right).
More mobile species are more vulnerable to targets and make a slower recovery,
all other things being equal.} \label{eradicationB}
   \end{center}
\end{figure}

\subsubsection*{`Siphoning' Out the Reserve Population from the Boundary}

\begin{figure}[H]
    \begin{center}
\includegraphics[width=7.7cm, angle=0, clip = true]{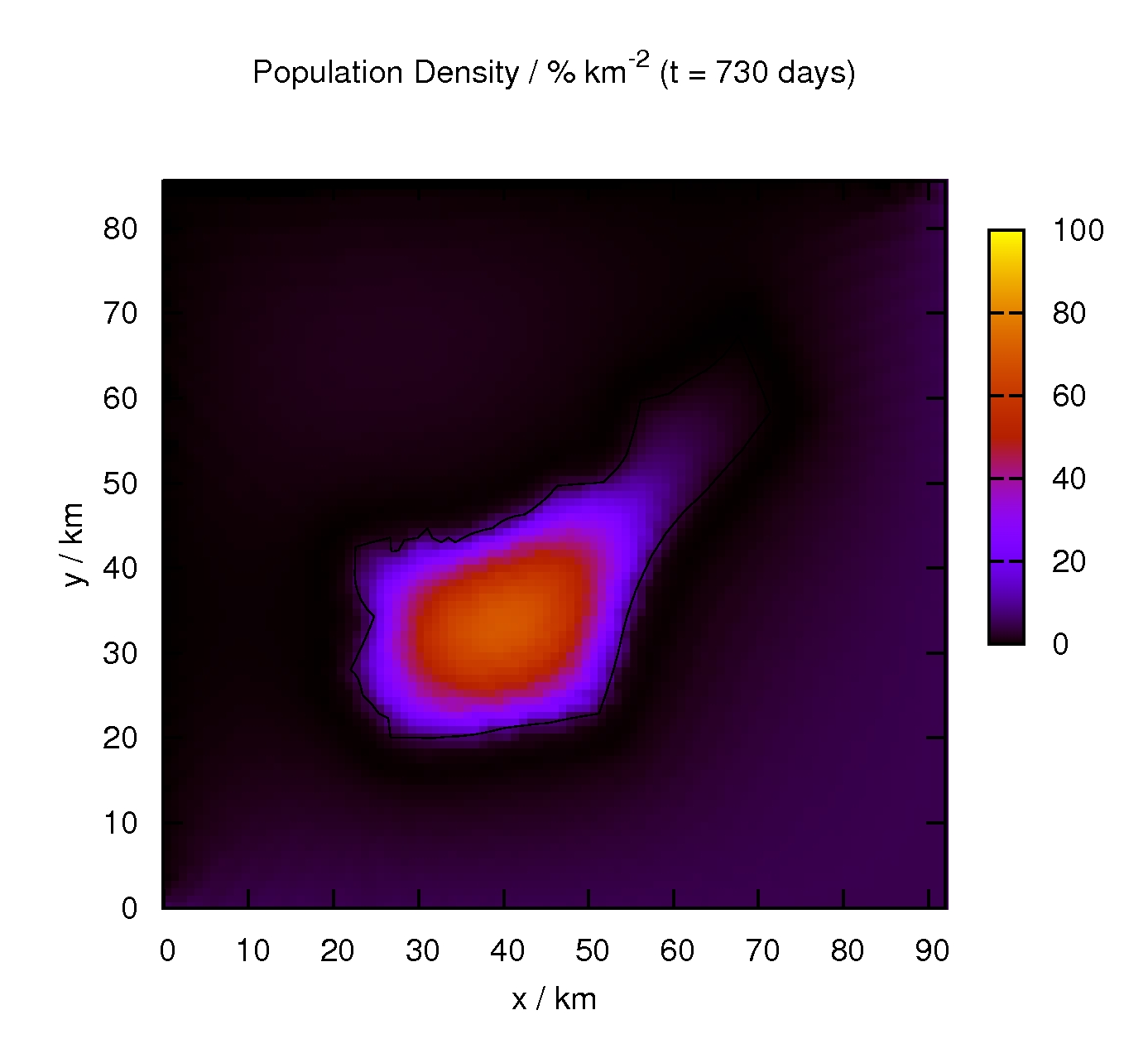}
\caption{The effect of a 5 $\mathrm{km}$-wide barrier in which a 50 \%
$\mathrm{day}^{-1}$ mortality is imposed throughout, for a period of 2 years,
using $\lambda = 0.32 \ \mathrm{km}^2 \ \mathrm{day}^{-1}$. The reserve is 960
$\mathrm{km}^2$ and its tsetse populations cannot be removed from the boundary
using current target technology.} \label{pumping}
   \end{center}
\end{figure}

\section{Improvising Barriers According to Specified Mortality Rates}

Odour-baited targets and cattle treated with so-called `pour-ons' are the means by which tsetse barriers can be constructed.

\subsection{Odour-Baited Targets (and the Possible Revelation of Competition)}

The definitive experimental work involving target barriers for {\em G. austeni}
and {\em G. brevipalpis} is that of {\sc Esterhuizen}, {\sc Kappmeier Green},
{\sc Nevill} and {\sc Van Den Bossche}
\cite{EsterhuizenKappmeierGreenNevillVanDenBossche}. Essentially they barricaded
a small peninsula with targets. They also placed targets on the peninsula itself
and measured the decline of {\em G. austeni} and {\em G. brevipalpis} in
relation to the target density on the peninsula. At a target density of 8
$\mathrm{km}^{-2}$ {\em G. austeni} was found to decline at a growth rate of
around - 0.014 $\mathrm{day}^{-1}$. 

The translation of such a growth rate into target mortality is facilitated
by {\sc Williams}, {\sc Dransfield} and {\sc Brightwell}
\cite{Williams1}'s seminal equation. In their equation
\begin{eqnarray} \label{6}
\beta e^{ - \tau_{0}(\delta_{0} + R) - \tau_{1}(\delta_{1} + R) -
\tau_{2}(\delta_{2} + R)} &=& 1 - e^{ - \tau_{2}(\delta_{2}
+ R)},
\end{eqnarray}
$R$ is the growth rate, $\beta$ is the fecundity, the $\delta_i$ are
mortalities, the $\tau_i$ are durations and the subscripts 0, 1, and 2 pertain
to the puparial, pre-ovulatory and interlarval stages respectively. Despite good
estimates of the rate of {\em G. austeni} decline, choosing other parameters in
the equation is something of a heuristic exercise. If one takes cognizance of
{\sc Rogers} and {\sc Randolph} \cite{RogersAndRandolph1}'s findings on
predation, the pupal water loss model of {\sc Childs} \cite{Childs2} etc., a 2
\% $\mathrm{day}^{-1}$ natural mortality rate among pupae would not be
unreasonable for the species in question. Natural mortality for the pre-ovulatory
cohort is known to be high. It includes massive teneral mortality and although
it does not fall victim to targets in the same proportions as the adults, some
still do ({\sc Hargrove} \cite{Hargrove9}). With this fact in mind the
pre-ovulatory stage flies were assigned a natural mortality of 2.2 \%
$\mathrm{day}^{-1}$ and assumed to have a target-related mortality one half
that of adults. At 22.1$^\circ$C the relevant
formulae\footnotemark[1]\footnotetext[1]{{\sc Hargrove} \cite{Hargrove3}'s
improved {\sc East African High Commision} \cite{Anonymous} formulae.} for the
first and subsequent interlarval periods predict 17.5 days and 10.5 days
respectively. The formula Equation \ref{2} gives a puparial duration of 34.6
days for females. A miscarriage rate of 5\%, and therefore a fecundity of 0.475
was used (in keeping with {\sc Williams et al.} \cite{Williams1}). Solving
Equation \ref{6}, using Newton's method and assuming a pre-existing equilibrium
involving the aforementioned parameters, suggested a natural mortality of 1.77
\% $\mathrm{day}^{-1}$ among the adults.

Solving Equation \ref{6}, using Newton's method and the newly completed set of 
parameters, suggested the 8 $\mathrm{km}^{-2}$ target density of {\sc
Esterhuizen et. al.} \cite{EsterhuizenKappmeierGreenNevillVanDenBossche} was
equivalent to an artificially imposed mortality of 2.39 \% $\mathrm{day}^{-1}$
(0.30 \% $\mathrm{day}^{-1}$ per target). Target-related mortality is obviously
much lower for these forest species. In comparison, a single odour-baited
target\footnotemark[2]\footnotetext[2]{As specified in {\sc Vale, Hargrove,
Cockbill} and {\sc Phelps} \cite{ValeHargroveCockbillAndPhelps}.} kills 2 \%
$\mathrm{day}^{-1}$ of the female {\em G. pallidipes} population ({\sc Hargrove}
\cite{Hargrove6}). 

{\sc Remark:} Note that a value of 2.39 \% $\mathrm{day}^{-1}$ is in very close
agreement with the modelled, 2 \% $\mathrm{day}^{-1}$ barrier-mortality
required for the isolation of {\em G. austeni}, as are the widths of the
modelled barrier and {\sc Esterhuizen et al.}
\cite{EsterhuizenKappmeierGreenNevillVanDenBossche}'s peninsula barricade
respectively. The work of {\sc Esterhuizen et al.}
\cite{EsterhuizenKappmeierGreenNevillVanDenBossche} was only subsequently
brought to the author's attention. 

So far as {\em G. brevipalpis} is concerned, {\sc Esterhuizen et al.}
\cite{EsterhuizenKappmeierGreenNevillVanDenBossche}'s results are not as clear.
Indeed, the results of this work suggest their barrier-zone might have been
completely ineffectual against a very mobile {\em G. brevipalpis}. Another
possibility is that {\em G. brevipalpis} is completely impartial to
odour-baited targets. Yet something certainly did happen in both
target-containing sectors when the concentration of targets reached a density
of 8 $\mathrm{km}^{-2}$. {\em G. brevipalpis} initially declined at a rate
indicative of an imposed mortality of 0.63 \% $\mathrm{day}^{-1}$ (using
Equation \ref{6}, again). A subsequent reversal of this decline then coincided
with the demise of {\em G. austeni} and the {\em G. brevipalpis} population grew
to levels never previously attained; in spite of the targets. A number of
explanations spring to mind. One argument is that the data is too poor, that
what is being observed is simply random noise, should be ignored. Another
possibility is that there was a delay in recolonization by this highly mobile
species. Certainly one has good reason to suspect an element of diffusion to be
operative, even if not the overriding analysis. Why then, the delay? There could
be reasons. 

An alternative explanation is that the reversal in fortune of the one species
coincided with the demise of the other due to the two being in competition: So
deleterious was the presence of {\em G. austeni} to {\em G. brevipalpis} that
its removal is able to counteract the imposition of a 0.63 \%
$\mathrm{day}^{-1}$ mortality on {\em G. brevipalpis} (a decline in growth rate
of - 0.0039 $\mathrm{day}^{-1}$). In retrospect, such a situation might have
been anticipated. Indeed, one of the posits of this model is that the
limitations on growth at pupal sites are density dependent. Pupal habitat for
{\em G. brevipalpis} is more stringently confined than for {\em G. austeni}
(according to demonstrations of the pupal water loss model in {\sc Childs}
\cite{Childs2}) and the {\em G. austeni} puparial duration is a full 20\%
shorter than that of {\em G. brevipalpis} ({\sc Parker} \cite{Parker1}). One
would imagine {\em G. brevipalpis} also has an adverse effect on {\em G.
austeni}. Just how severe and whether or not it can be exploited, is not
evident. Further experimentation is required. That {\sc Esterhuizen et al.}
\cite{EsterhuizenKappmeierGreenNevillVanDenBossche} were simply not able to
measure a true value for the target-related mortality of {\em G. brevipalpis},
owing to high diffusion rates, is their own conclusion. 

If the accepted wisdom is correct that the effect of uniformly distributed
targets is additive, then a given mortality may be designed in terms of Table
\ref{targetMortality} as follows. 
\begin{table}[H]
    \begin{center}
\begin{tabular}{c | c c c c}  
&  &  &  & \\
$\delta$ / $\mathrm{day}^{-1}$ \hspace{5mm} & \hspace{5mm} 0.02 & \hspace{5mm} 0.04 & \hspace{5mm} 0.08 & \hspace{5mm} 0.12 \\ 
&  &  &  & \\ \hline 
&  &  &  & \\
{\em G. austeni} \hspace{5mm} & \hspace{5mm} 7 & \hspace{5mm} 13 & \hspace{5mm} 27 & \hspace{5mm} 40 \\ 
&  &  &  & \\
{\em G. brevipalpis} \hspace{5mm} & \hspace{5mm} 25 & \hspace{5mm} 51 & \hspace{5mm} 102 & \hspace{5mm} 152 \\ 
&  &  &  & \\
{\em G. pallidipes} \hspace{5mm} & \hspace{5mm} 1 & \hspace{5mm} 2 & \hspace{5mm} 4 & \hspace{5mm} 6 \\ 
&  &  &  & \\
\end{tabular}
\caption{The number of targets per $\mathrm{km}^2$ which will produce a required daily mortality, $\delta$, for each species.} \label{targetMortality}
    \end{center}
\end{table}

\subsection{Tethered, Treated Cattle}

Unpublished experiments by S. J. Torr (reported in {\sc Hargrove}, {\sc Torr}
and {\sc Kindness} \cite{HargroveTorrAndKindness}) suggest that a single
odour-baited target kills the equivalent number of {\em G. pallidipes} females
in 1 $\mathrm{km}^2$ as an insecticide-treated ox of weight 400 $\mathrm{kg}$
does in a day. Since  {\sc Esterhuizen et al.}
\cite{EsterhuizenKappmeierGreenNevillVanDenBossche} used the same 1.5 $\times$ 1
m, black-blue-black targets (manufactured by Bonar Industries,
Harare)\footnotemark[1]\footnotetext[1]{{\sc Esterhuizen} \cite{Esterhuizen1}
and {\sc Hargrove} \cite{Hargrove9}}, the corresponding target-related
mortality should apply to the ox for {\em G. austeni} and {\em G. brevipalpis};
assuming these species do not discriminate any differently between the chemical
signatures of the beast and the target. 

\subsection{Treated Herds}

In {\sc Hargrove}, {\sc Holloway}, {\sc Vale}, {\sc Gough} and {\sc Hall}
\cite{HargroveHollowayValeGoughAndHall} it was determined that tsetse catches
changed with the tonnage of cattle, $m$, in a ventilated shed and could be
described by
\begin{eqnarray*}
\delta \propto 4 m^{0.475}.
\end{eqnarray*}
Torr's experiment (reported in {\sc Hargrove et al.}
\cite{HargroveTorrAndKindness}) allows the constants in the simplistic model,
\begin{eqnarray*}
\left[ \begin{array}{c} 0.0030 \\ 0.00079 \\ 0.02 \end{array} \right] &=& 4 \ \left[ \begin{array}{c} c_{\scriptsize \mbox{austeni}} \\ c_{\scriptsize \mbox{brevipalpis}} \\ c_{\scriptsize \mbox{pallidipes}} \end{array} \right] \ 0.4^{0.475}, 
\end{eqnarray*}
to be determined. The minimum tonnage of cattle required to induce a given daily mortality in a square kilometre is therefore given by Table \ref{tonnage}.
\begin{table}[H]
    \begin{center}
\begin{tabular}{c | c c c }
&  &  & \\
species & {\em G. austeni} & {\em G. brevipalpis} & {\em G. pallidipes} \\ 
&  &  & \\ \hline 
&  &  & \\ 
herd mass / tons $\mathrm{km}^{-2}$ & $\displaystyle 0.4 \left( \frac{\delta}{0.0030} \right)^{\frac{1}{0.475}}$ & $\displaystyle 0.4 \left( \frac{\delta}{0.00079} \right)^{\frac{1}{0.475}}$ & $\displaystyle 0.4 \left( \frac{\delta}{0.02} \right)^{\frac{1}{0.475}}$ \\
&  &  & \\ 
\end{tabular}
\caption{The treated herd mass required to bring about a given mortality, $\delta$, in each species.} \label{tonnage}
    \end{center}
\end{table}

\section{Conclusions}

The diffusion coefficient for {\em G. austeni} is probably around 0.04
$\mathrm{km}^2$ $\mathrm{day}^{-1}$, although a worst-case value of 0.08
$\mathrm{km}^2$ $\mathrm{day}^{-1}$ was assumed (Figures \ref{risk} and
\ref{1st}). The diffusion coefficient for {\em G. brevipalpis} can be assumed to
be around 0.32 $\mathrm{km}^2$ $\mathrm{day}^{-1}$, although it could be as low
as 0.16 $\mathrm{km}^2$ $\mathrm{day}^{-1}$ (Figures \ref{risk} and \ref{2nd}). 

Based on the worst-case values in terms of which the problem was phrased, the
simulations suggest that the temporary imposition of a 2 \% $\mathrm{day}^{-1}$
mortality everywhere outside the reserve for a period of 2 years will have no
lasting effect on the influence of the reserve when it comes to either
population; although it certainly will eradicate tsetse from areas of poor
habitat, outside the reserve (Figures \ref{pourOnsA1} and \ref{pourOnsB1}). It
is doubtful whether the populations within the reserve can be `siphoned' or
`pumped out' to extinction, or even the 20\% level, from outside the reserve
boundary (Figure \ref{pumping}).

During the initial stages of this work, it became apparent that 2.5
$\mathrm{km}$-wide, target barriers were not efficacious (Figure \ref{2.5km})
and their further investigation was abandoned in favour of a 5 $\mathrm{km}$ 
width (Figure \ref{5kmA}). The influence of the reserve on surrounding {\em G.
austeni} population levels can be completely neutralized by a 5
$\mathrm{km}$-wide barrier in which there is a mortality of 2 \%
$\mathrm{day}^{-1}$ throughout (Figure \ref{impenetrableA}). For {\em G.
brevipalpis} a 5 $\mathrm{km}$ wide barrier to the same end will require a
mortality of 8 \% $\mathrm{day}^{-1}$ throughout (Figure \ref{impenetrableB}).
Notice, however, that these measures are not in any way able to address the
likelihood of a greater prevalence, as well as more lethal strains, of
trypanosome infection in flies close to the reserve boundary, regardless of any
reduction in their numbers. 

A 5 $\mathrm{km}$-wide barrier of odour-baited targets with a mortality of 4
\% $\mathrm{day}^{-1}$, throughout, should succeed in isolating a worst-case, 
Hluhluwe-iMfolozi {\em G. austeni} population and its associated
trypanosomiasis from the surrounding areas (Figures \ref{5kmA} and
\ref{impenetrableA}). A more optimistic estimate of its mobility suggests a
mortality of 2 \% $\mathrm{day}^{-1}$ will suffice (Figure \ref{5kmA}). These
mortalities correspond to a deployment of odour-baited
targets\footnotemark[1]\footnotetext[1]{As used by {\sc Esterhuizen et al.}
\cite{EsterhuizenKappmeierGreenNevillVanDenBossche}.} with a minimum density of
13 $\mathrm{km}^{-2}$ and 7 $\mathrm{km}^{-2}$ respectively. Simple arguments
suggest that such counter measures should reduce the {\em G. austeni} problem
associated with the reserve by at least an order of magnitude. 

For {\em G. brevipalpis}, a mortality of 12 \% $\mathrm{day}^{-1}$, throughout,
will achieve the same end of complete isolation (Figure \ref{impenetrableB}).
The impartiallity of {\em G. brevipalpis} to odour-baited targets is obviously
a concern, should this species be conclusively shown to be a vector of
trypanosomiasis. A 12 \%  $\mathrm{day}^{-1}$ mortality is not practical in
terms of what one can only surmise is the mortality of current odour-baited
target technology. A mortality of 8 \% $\mathrm{day}^{-1}$ fails mainly from the
point of view of the crude sensitivity analysis and it will probably suffice if:
the width of the barrier can, with absolute certainty, be said to be no less
than 5km; boundaries are non-concave; the context is one involving rebound
suppression, following the use of pour-ons for a 2 year period. The less
ambitious goal of neutralizing the reserve's influence on the surrounding {\em
G. brevipalpis} population would require a deployment of odour-baited targets
with a minimum density of 102 $\mathrm{km}^{-2}$; again a clearly impractical
proposition. Any strategy for the control of {\em G. brevipalpis} should include
the surroundings of the Hluhluwe dam and its backwater, as if it were part of
the reserve, based on {\sc Childs} \cite{Childs2}. 

Extrapolating the work of {\sc Hargrove et al.}
\cite{HargroveHollowayValeGoughAndHall} suggests that the substitution of
insecticide-treated herds for odour-baited targets is not a viable alternative
for the control of the two forest species in question. Required herd-masses are
impractical for the purposes of barriers and containment. Individually tetherd,
treated cattle can be used as substitutes for odour-baited targets, although
the numbers required are probably not really practical either. Periodically
rotating them in and out of the barrier zone would prevent a loss of resistance
to tick-bourne diseases and an enzootic condition. (Individually tethered,
deltamethrin-treated cattle, distributed uniformly throughout a barrier zone,
may be less likely than targets to fall victim to the tragedy of the commons
type mentality known to prevail among the local population.) 

The premise that the entire reserve, and it alone, is a problem is not as valid
for {\em G. austeni} as it is for {\em G. brevipalpis} (Figure \ref{risk}). In
the case of {\em G. austeni} it may well be worth singling out individual locii
for the application of control measures (e.g. the flood plain of the Hluhluwe
River, the vicinities of the Hluhluwe Dam, its backwater and the confluence of
the Black Mfolozi and White Mfolozi rivers), based on {\sc Childs}
\cite{Childs2} and Figure \ref{risk}. For {\em G. brevipalpis}, however, a
comparison of Figures \ref{risk} and \ref{2nd} suggests that the logistic growth
rate might even exceed 1.7 \% $\mathrm{day}^{-1}$ in the northern, Hluhluwe
sector of the reserve. So favourable is that habitat. 

For a given mortality, more mobile species are found to be more vulnerable to
eradication than more sedentary species while the opposite is true for
containment. The scenarios depicted in Figures \ref{eradicationA} and
\ref{eradicationB} demonstrate, firstly, that high diffusion rates are more
amenable to eradication since the same target consistancy is not required
locally. Species with high diffusion rates are vulnerable to controls which are
geographically more remote. Secondly, high diffusion rates lead to a much weaker
initial recovery from levels close to
extinction\footnotemark[2]\footnotetext[2]{Although this could be an artefact of
assuming more mobile species have the same growth rate as more sedentary
species.}. The reason is that there is a tendency to disperse which is not
efficacious for logistic growth at low population densities. 

One is now presented with a scenario in which {\em G. brevipalpis} may be more
vulnerable to eradication than containment and vice versa for {\em G. austeni}
(partiality to odour-baited targets and existing population levels aside). Yet
whether or not {\em G. brevipalpis} is even an agent of infection is still a
moot point ({\sc Motloang et al.}
\cite{MotloangMasumuVanDenBosscheMajiwaLatif}). {\em G. austeni} is, in
contrast, without the slightest doubt a highly competent vector of
trypanosomiasis. 
The possibility that eliminating {\em G. brevipalpis} will create further
opportunity for {\em G. austeni} and, consequently, trypanosomiasis needs to be
considered. The experimental results of {\sc Esterhuizen et al.}
\cite{EsterhuizenKappmeierGreenNevillVanDenBossche} can be interpreted to lend
credence to exactly such a theory. They could suggest intense competition
between the two species, to the extent that {\em G. brevipalpis} may actually
benefit from odour-baited targets should their density be sufficient to
eliminate {\em G. austeni} only. The existance of a reciprocal effect on {\em G.
austeni} may be worth investigating . It could be exploitable. Then again, what
is observed could simply be a delayed invasion response or even random noise.
That {\sc Esterhuizen et al.}
\cite{EsterhuizenKappmeierGreenNevillVanDenBossche} were simply not able to
measure the target-related mortality of {\em G. brevipalpis}, owing to high
diffusion rates, is an alternative conclusion. 

The K. P. P. equation can be solved by way of the application of the finite
element method for the spatial discretisation and a backward difference for the
temporal discretisation. This same strategy augmented by the linearisation and
iteration of the nonlinear term also worked well for Fisher's equation, with
good convergence for the range of conditions investigated. There is a certain
amount of academic interest in this more challenging mathematics in that, if
the numerical techniques employed are powerful enough to solve a nonlinear
Fisher's equation, they will, logically, solve an equation with any other
variants of the logistic term contemplated. This has important implications for the modelling of other vector-bourne diseases. 

A comparison of the results of the two models offered some interesting insights.
One concern at the outset was that if the age profile is altered in such a way
that it contains a significantly higher proportion of pre-ovulatory flies, then
the logistic growth rate (which is based on a fixed age profile) may no longer
be appropriate. Just how reasonable is the assumption of a fixed age profile?
Fisher's equation makes a far worse assumption in that it not only denies the
existance of the pre-ovulatory stage, it also denies the existance of the entire
pupal phase. A comparison between results of the two models might indicate the
extent of the problem. The combination of warm temperatures, low imposed
mortalities and long two year cycles gave the population ample time to
re-equilibrate in this particular case study, with the result that there was no
discernable difference between the K.P.P.-model results and those arising from
the unquestioning application of Fisher's equation (although this was not
necessarily the case at lower temperatures). The suggestion is, therefore, that
the assumption of a fixed age profile (in light of the longer than usual first
interlarval period and artificially imposed mortalities) is permittable. If
circumstances permit the denial of the existance of the pupal phase (Fisher's
equation), then it stands to reason that a failure to recognize the existance of
the far shorter pre-ovulatory phase (K.P.P. and Fisher's equation) should be
permitted. Finally, convergence with little, or no iteration for Fisher's
equation was useful in suggesting the steady state. 

It is, nonetheless, inadviseable to use a model based on an unmodified Fisher's
equation for tsetse. Lower temperatures or catastrophic mortalities inflicted,
for example, by an aerial spray, are all circumstances in which attributing
subsequent growth to a current, as opposed to historical, population would be
profoundly incorrect. Notice that the model based on historical parentage would,
under the latter circumstance, still fail to take the subsequent reproductive
phase entrainment and altered age profile into account. Reproductive rates would
initially be over-estimated, later, under-estimated and so on. Unlike Fisher's
equation, however, the model is expected to recover. So long as circumstances
allow the population to re-equilibrate there are unlikely to be any problems. 

\section{Acknowledgements} 

Abdalla Latif and the Onderstepoort Veterinary Institute are thanked for their
generousity in both facilitating and funding this research. Guy Hendrickx is
gratefully acknowledged for donating the two maps of tsetse risk and Andrew
Parker is thanked for the information on puparial durations. Other, general
information on the Hluhluwe-iMfolozi game reserve was supplied by Ezemvelo
K.Z.N. Wildlife and the satellite image was kindly supplied by Marina Faber of
Peace Parks Foundation. Brian Williams is thanked for taking a general interest
in this work and John Hargrove, for sharing his vast knowledge of tsetse. This
work obviously relies on the monumental research efforts and pioneering work
carried out by the cited authors.  

\nocite{Murray1}

\nocite{Anonymous2}


\bibliography{kznReservesPaper}

\section*{Addendum}

\subsection*{The Change in Population Density Due to Migration}

\renewcommand{\thefootnote}{\fnsymbol{footnote}}
Consider the hypothetical scenario of a mobile population in the absence of
either reproduction or mortality (external and artificially imposed, or
otherwise). Let $\Omega(t)$ be an arbitrary sub-volume of flies with boundary
$\Gamma(t)$. Then biomass should be conserved so that \footnotetext[2]{The
material derivative of the Jacobian is given by the kinematic relation ${\dot
J}_0 = J_0 \mathop{\rm div}{\mathbf{v}}$, a result demonstrated in most popular
textbooks on continuum mechanics (eg. {\sc Marsden} and {\sc Hughes}
\cite{mh:1}).} 
\begin{eqnarray} \label{92}
\frac{D}{Dt} \int_{\Omega(t)} \rho \ d\Omega &=& 0 \hspace{10mm}
\mbox{(rate of change of mass with time }= \ 0\mbox{)}
\nonumber \\ 
& & \nonumber \\
\frac{d}{dt} \int_{\Omega_0} {{\rho}_0} J_0 \ {d \Omega_0} &=& 0
\hspace{10mm} \mbox{(reformulating in the material}
\nonumber \\
& & \hspace{13mm} \mbox{configuration, } \Omega_0 \mbox{)}
\nonumber \\
\int_{\Omega_0} \frac{d}{dt}  \left\{ {{\rho}_0} J_0 \right\} \ d
\Omega_0 &=& 0 \hspace{10mm} \mbox{(since limits are not time
dependent} \nonumber \\
& & \hspace{13mm} \mbox{in the material configuration)} \nonumber \\
\int_{\Omega_0} \left( \rho_0 {\dot J}_0 \ + \ {\dot \rho}_0
J_0 \right) \ d \Omega_0 &=& 0 \hspace{10mm} \mbox{(by the
product rule)} \nonumber \\
& & \nonumber \\
\int_{\Omega(t)} \left( \dot{\rho} \ + \ {\rho \mathop{\rm div}{\mathbf{v}} }
\right) \ d\Omega &=& 0 \hspace{10mm} \mbox{(using } \dot{J_0} = J_0 \mbox{div}\, {\mathbf{v}} \mbox{)\footnotemark[2]}, \nonumber 
\end{eqnarray}
in which $\mathbf{v}$ is velocity. Since the volume was arbitrary it
follows that the integrand must be zero. That is
\begin{eqnarray*}
\dot{\rho} \ + \ {\rho \mathop{\rm div}{\mathbf{v}} } &=& 0. 
\end{eqnarray*} 
Now consider this biomass conservation in the context of another arbitrary sub-volume, this time of habitat, $\Omega_h$, with boundary $\Gamma_h$. 
\begin{eqnarray}
\int_{\Omega_h} \left( \frac{\partial \rho}{\partial t} \ + \ \nabla \rho \cdot \mathbf{v} \ + \ {\rho \mathop{\rm div}{\mathbf{v}} }
\right) \ d\Omega_h &=& 0 \hspace{10mm} \mbox{(expanding }  \dot{\rho} \mbox{)} \nonumber \\
& & \nonumber \\
\int_{\Omega_h} \left( \frac{\partial \rho}{\partial t} \ + \ \mathop{\rm div} \{ \rho \mathbf{v} \} \right) \ d\Omega_h &=& 0 \hspace{10mm} \mbox{(by the
product rule)} \nonumber \\
& & \nonumber \\
\int_{\Omega_h} \frac{\partial \rho}{\partial t} \ d\Omega_h \ + \ \int_{\Gamma_h} \rho \ \mathbf{v} \cdot \mathbf{n} \ d\Gamma_h &=& 0 \hspace{10mm} \mbox{(by the divergence theorem)} \nonumber \\
& & \nonumber \\
\int_{\Omega_h} \frac{\partial \rho}{\partial t} \ d\Omega_h - \ \int_{\Gamma_h} - \lambda \ \nabla \rho \cdot (-\mathbf{n}) \ d\Gamma_h &=& 0 \hspace{10mm} \mbox{(by Fick's 1st law)} \nonumber \\
& & \nonumber \\
\int_{\Omega_h} \left( \frac{\partial \rho}{\partial t} \ - \lambda \mathop{\rm div} \nabla \rho \right) \ d\Omega_h &=& 0 \hspace{10mm} \mbox{(by the divergence theorem).} \nonumber
\end{eqnarray} 
Since the volume was arbitrary it again follows that the integrand must be
zero. That is
\begin{eqnarray*}
\frac{\partial \rho}{\partial t} &=& \lambda \ \mathop{\rm div} \nabla \rho,
\end{eqnarray*} 
in which $\lambda$ is the diffusion coefficient.  
\renewcommand{\thefootnote}{\arabic{footnote}}

\subsection*{Dimensionless Form}

Equation \ref{1} is converted to its dimensionless form, as is standard practice
before commencing a computation of this nature. (One wouldn't want the solution
to be influenced in any way by the choice of units.) Suppose that $T$ is (only
for the present) an unit of time, $X$ is a unit of length and $\eta$ is an unit of population density. The scaled variables are then
\begin{equation*}
{\mathbf{x}} = { \bar {\mathbf{x}} }X, \hspace{5mm} t = {\bar t } T \hspace{5mm} \mbox{and} \hspace{5mm} \rho = {\bar \rho} \eta \mbox{ \ \ (including }K = {\bar K} \eta \mbox{).} 
\end{equation*}
Thus Equation \ref{1} can be rewritten as
\begin{eqnarray*}
\frac{\eta}{T} \frac{\partial {\bar \rho}}{\partial {\bar t}} &=&
\frac{\eta}{X^2} \ \lambda \ \mathop{\bar {\rm div}} {\bar \nabla} {\bar \rho} +
r \eta {\bar \rho}^* \left( 1 - \frac{ {\bar \rho}^* }{\bar K} \right) - \delta
\eta {\bar \rho}.
\end{eqnarray*}
All of this suggests using $T = \displaystyle \frac{X^2}{\lambda}$ and $X = \displaystyle \sqrt{\frac{\lambda}{r}}$ so
that the above equation becomes
\begin{eqnarray*} \label{20}
\frac{\partial {\bar \rho}}{\partial {\bar t}} &=& \mathop{\bar {\rm div}} {\bar
\nabla} {\bar \rho} + {\bar \rho}^* \left( 1 - \frac{\bar \rho^*}{\bar K} 
\right) - \frac{ \delta }{r} {\bar \rho}.
\end{eqnarray*}
If the mesh is in units of kilometres, for example, then it must be converted by
dividing through by $\sqrt{\frac{\lambda}{r}}$ kilometres. 

{\bf Complication:} If one intends accomodating any environmental variation in the rates of diffusion and growth, the scaled equation will entail different and therefore irreconcileable time steps. The equation
\begin{eqnarray} \label{10}
\frac{\partial {\bar \rho}}{\partial {\bar t}} &=&
\frac{\lambda}{\lambda_{\scriptsize scale}} \mathop{\bar {\rm div}} {\bar
\nabla} {\bar \rho} + \frac{r}{r_{\scriptsize scale}}{\bar \rho}^* \left( 1 -
\frac{\bar \rho^*}{\bar K} \right) - \frac{ \delta }{r_{\scriptsize scale}} {\bar \rho}
\end{eqnarray}
allows a time discretisation which conforms.

\subsection*{Variational Formulation} 

A variational formulation of Equation \ref{10} is obtained in the usual fashion;
premultiplying the primitive variable equation by an arbitrary function, $w$,
and integrating over the domain, $\Omega$, gives rise to the equation
\begin{eqnarray*} 
\int_{ \Omega } w \ \frac{\partial \rho}{\partial t} \ {d{\Omega}}  &=&
\frac{\lambda}{\lambda_{\scriptsize scale}} \int_{ \Omega } w \ \mathop{\rm
div}{{\nabla} {\rho }} \ {d {\Omega}} + \frac{r}{r_{\scriptsize scale}} \int_{
\Omega } w \ \rho^* \left( 1 - \frac{ \rho^* }{K} \right) \ {d {\Omega}} -
\frac{ \delta }{r_{\scriptsize scale}} \int_{ \Omega } w \ \rho \ {d {\Omega}}.
\end{eqnarray*}
The approximation-wise cumbersome second derivatives can also be avoided in the
usual fashion. The term which contains the divergence of ${\nabla} {\rho }$ can
be regarded as one part of a differentiated product and the divergence theorem
applied so that
\begin{eqnarray*}
\int_{ \Omega } w \frac{\partial {
\rho}_{,i}}{\partial x_i} \ {d {\Omega}} &=& \int_{ \Gamma}
w {\rho}_{,i} {n}_i \ d{\Gamma} - \int_{
\Omega } {w}_{,i} {\rho}_{,i} \ {d {\Omega}},
\end{eqnarray*}
in which $\mathbf{n}$ is the outward unit normal and $\Gamma$ is the domain
boundary. The boundary integral obviously vanishes for a von Neumann,
$\mathbf{n} \cdot \nabla \rho = 0$ type boundary condition while the arbitrary
vector of the formulation can be assigned a value of zero where boundary
conditions are Dirichlet (and an equation is consequently not required). The
boundary integral is therefore irrelevant.

\end{document}